\pdfoutput=1

\documentclass[utf8]{frontiersSCNS}

\usepackage{url,lineno,microtype,subcaption}
\usepackage[onehalfspacing]{setspace}

\newcommand{\arcsec}{{\hbox{$^{\prime\prime}$}}}
\newcommand{\degr}{{\hbox{$^\circ$}}}

\def\keyFont{\fontsize{8}{11}\helveticabold }
\def\firstAuthorLast{Nindos} 
\def\Authors{Alexander Nindos\,$^{1,*}$}
\def\Address{$^{1}$Physics Department, University of Ioannina, Ioannina GR-45110, Greece }
\def\corrAuthor{A. Nindos}

\def\corrEmail{anindos@uoi.gr}

\begin{document}
\onecolumn
\firstpage{1}

\title[Incoherent Solar Radio Emission]{Incoherent Solar Radio Emission} 

\author[\firstAuthorLast ]{\Authors}
\address{} 
\correspondance{} 

\extraAuth{}

\maketitle

\begin{abstract}

\section{}
Incoherent solar radio radiation comes from the free-free,
gyroresonance, and gyrosynchrotron emission mechanisms. Free-free is
primarily produced from Coulomb collisions between thermal electrons
and ions. Gyroresonance and gyrosynchrotron result from the
acceleration of low-energy electrons and mildly relativistic
electrons, respectively, in the presence of a  magnetic field. In the
non-flaring Sun, free-free is the dominant emission mechanism with the
exception of regions of strong magnetic fields which emit
gyroresonance at microwaves. Due to its ubiquitous presence,
free-free emission can be used to probe the non-flaring solar
atmosphere above  temperature minimum. Gyroresonance opacity depends
strongly on the magnetic  field strength and orientation; hence it
provides a unique tool for the  estimation of coronal magnetic
fields. Gyrosynchrotron is the primary emission mechanism in flares
at frequencies higher than 1-2 GHz and depends on  the
properties of both the magnetic field  and the accelerated electrons,
as well as the properties of the ambient plasma.  In this paper we
discuss in detail the above mechanisms and their 
diagnostic potential.

\tiny
 \keyFont{ \section{Keywords:} Sun, Solar radio emission, chromosphere, corona, quiet Sun, active regions, flares} 
\end{abstract}

\section{Introduction}

The Sun produces radiation across virtually the entire electromagnetic
spectrum. Radio frequencies offer valuable diagnostic potential
because two of the natural frequencies of the atmosphere of the Sun,
the electron plasma frequency and the electron gyrofrequency, belong
to the radio band.

In the Sun there are no significant spectral lines,  either in
emission or absorption, from millimeter to meter wavelengths
\citep[but see][for a possible detection]{Dravskikh88}; pressure
broadening is so high that makes such lines undetectable.  In the Sun
the free electrons dominate the radio emission mechanisms. Solar radio
emission is produced from electrons with either a thermal or a
nonthermal distribution, and the emission can be either incoherent or
coherent.  In incoherent mechanisms, no back-reaction of the emission
on the electron distribution is present, and the emitted photons show
no phase association  while their number is proportional to the number
of electrons. In coherent mechanisms, all electrons exhibit
acceleration in phase; they act together to generate photons that are
in phase.

Coherent radiation due to wave-particle and wave-wave interactions
plays an important role in transient phenomena at frequencies below 1-2 GHz.  
Coherent emission mechanisms are discussed elsewhere in this issue by 
Fleishman.
   
There are two classes of incoherent emission mechanisms that are
important on the Sun: free-free (or bremsstrahlung) and
gyromagnetic. At radio frequencies free-free emission is primarily
produced from collisions between ions and thermal electrons and
dominates the radio emission of the quiet Sun. Furthermore it
contributes significantly to the radio emission of non-flaring active
regions and of certain flares during their decay phase. Erupting
prominences and coronal mass ejections  (CMEs) may also produce
free-free emission. Gyromagnetic radiation is produced from electrons
that are accelerated in the presence of magnetic fields. It is called
gyroresonance emission when it is produced by thermal  electrons with
energies that correspond to temperatures of the  non-flaring  corona
(about 10$^6$ K). Gyroresonance plays an important role in the
emission  above sunspots at microwaves. Gyromagnetic emission is
called gyrosynchrotron when it is produced by mildly relativistic
electrons having either nonthermal or thermal electron energy
distributions. Gyrosynchrotron is the principle incoherent emission
mechanism in flares.

There are several textbooks and review articles devoted to incoherent
solar radio emission. The classical textbooks by \cite{Kundu65},
\cite{Zheleznyakov70}, and \cite{Kruger79} are valuable sources of
information.  More recent textbooks include the volume edited by
D. Gary and  C. Keller \citep{Gary04}, as well as Aschwanden's
\citep{Aschwanden04}  book on the solar corona.  A historical account
of solar radio astronomy together with recent  developments has
been given in the review by \cite{Pick08}.  A review about the
radio emission of the quiet Sun has been published by
\cite{Shibasaki11} while reviews about gyroresonance have been
provided by \cite{White97} and  \cite{Lee07}. Transient incoherent
solar radio emission has been discussed by \cite{Bastian98} and
\cite{Nindos08}. A more recent book about incoherent microwave
emission from flaring loops has been written by \cite{Huang18}.

The structure of this paper is as follows. In \S2, we give a short
introduction on radiative transfer and propagation of radio emission.
In \S3, we discuss free-free emission, and in \S4 the different
types of gyromagnetic emission are outlined briefly. \S5 and \S6 are 
devoted to gyroresonance and gyrosynchrotron radiation, respectively. We 
present concluding remarks in \S7.

\section{Radiative transfer and propagation of radio emission}

\subsection{Radiative transfer basics}

Radiation is always intimately related to material through emission
and absorption processes. When both emission and absorption  are
considered,   the intensity, $I_{\nu}$, inside a plasma slab of
thickness  $dl$ changes by 

\begin{equation}
\frac{dI_{\nu}}{dl} = j_{\nu} \rho - k_{\nu} \rho I_{\nu}
\end{equation}
where $j_{\nu}$ and $k_{\nu}$ are the emission and absorption coefficients,
respectively, which are defined by
\begin{equation}
dI_{\nu} = j_{\nu} \rho dl
\end{equation}
and
\begin{equation}
dI_{\nu} = -k_{\nu} \rho I_{\nu} dl
\end{equation}
where $\rho$ is the plasma density. The above discussion is only valid
for thermal plasma radiation. For nonthermal electron distributions these
formulas hold if $\rho$ describes the concentration of nonthermal electrons.

It is convenient to discuss radiative 
transfer in terms of the optical depth, $\tau_{\nu}$, which is defined by:

\begin{equation}
d\tau_{\nu} = -k_{\nu} \rho dl
\end{equation}
Using the optical depth, eq. 1 becomes 

\begin{equation}
\frac{dI_{\nu}}{d\tau_{\nu}} = I_{\nu} - S_{\nu}
\end{equation}
where $S_{\nu}=j_{\nu}/k_{\nu}$ is the source function. Eq. 5 is called radiative
transfer equation (RTE). Its typical solution (i.e. the intensity at the 
observer where $\tau_{\nu}=0$) is

\begin{equation}
I_{\nu}(\tau_{\nu}=0) = \int_0^{\infty} S_{\nu} e^{-t_{\nu}} dt_{\nu}
\end{equation}
Therefore the intensity at the observer results from the contribution
of  all layers of a stellar atmosphere, with each layer contributing
proportionally to its emissivity, attenuated by the absorption of the
overlying layers,  $e^{-t_{\nu}}$.

From the RTE, we obtain for a finite slab of constant source function:

\begin{equation}
I_{\nu} = S_{\nu} (1 - e^{-\tau_{\nu}})
\end{equation}
In the optically thin case (i.e. transparent slab; $\tau_{\nu} \ll 1$), 
eq. 7 yields

\begin{equation}
I_{\nu} = \tau_{\nu} S_{\nu}
\end{equation}
while for the optically thick case (i.e. opaque slab; $\tau_{\nu} \gg 1$), 
we obtain

\begin{equation}
I_{\nu} = S_{\nu}
\end{equation}

Thermal solar radio emission originates from local thermodynamic
equilibrium (LTE) conditions (i.e. the temperature, $T$, does not
change much with respect to the mean free path of photons and free
electrons, and the anisotropy of the radiation field is small). In
LTE, the emission and absorption coefficients are not independent, but
the source function is equal to the Planck function. At radio
frequencies the inequality $h\nu/k_BT \ll 1$ holds, ($k_B$ is the Boltzmann 
constant), and the  Planckian simplifies to the Rayleigh-Jeans law. Then it is
convenient to define a brightness temperature, $T_b$, as the
equivalent temperature a black body would have in order to be as
bright as the observed brightness:

\begin{equation}
I_{\nu} = B_{\nu}(T_b) =\frac{2\nu^2}{c^2}k_B T_b
\end{equation}

Similarly, we can define an effective temperature, $T_{eff}$:

\begin{equation}
S_{\nu} = \frac{j_{\nu}}{k_{\nu}} = \frac{2\nu^2}{c^2} k_B T_{eff}
\end{equation}

Using our definitions of brightness temperature and effective temperature 
the RTE can be expressed as:

\begin{equation}
\frac{dT_b}{d\tau_{\nu}} = T_b - T_{eff}
\end{equation}

Similar to eq. 5, for an homogeneous source, the solution is:

\begin{equation}
T_b = T_{eff} (1 - e^{-\tau_{\nu}}), 
\end{equation}
For the optically thin and optically thick cases, eq. 13 yields

\begin{equation}
T_b = \tau_{\nu} T_{eff}
\end{equation}
and

\begin{equation}
T_b =  T_{eff}
\end{equation}
respectively. When the emission is incoherent, $T_{eff}$ is the
kinetic temperature in the case of thermal emission or corresponds to
the mean energy, $E$, of the emitting  electrons (i.e. $T_{eff}=E/k_B$)
when the emission is nonthermal. Therefore for an incoherent emission, $T_b$
is limited by the mean energy of the radiating particles. Since the
rest energy of the electron corresponds to $T_b=0.6 \times 10^{10}$ K,
we conclude that sources with $T_b \gg 10^{10}$ K cannot be due to
incoherent emission from non-relativistic or mildly relativistic
electrons. Incoherent emission by highly relativistic electrons is
dominated by synchrotron emission (see \S4), which is limited to $T_b
\leq 10^{12}$ K by Compton scattering \citep{Kellermann69}. 

\subsection{Propagation of radio emission}

In most cases, the corona can be described as a cold magnetized
plasma, and the magnetoionic theory \citep[e.g.][volume 1, chapter
  2]{Melrose80} is used to study the  propagating electromagnetic
modes. These are the  extraordinary ($x$-), ordinary ($o$-), $z$-, and
whistler mode. Only the $x$-  and $o$- mode waves can propagate from
the source to the observer, whereas the  $z$- and whistler mode waves
cannot due to stopbands in the refractive index.  For most
applications in solar radiophysics, the propagation of the $x$- and
$o$-mode waves can be described   by either the quasi-longitudinal
(QL) or the quasi-transverse (QT)  approximation (propagation almost
parallel and almost perpendicular to the magnetic field,
respectively). 

For observational purposes it is easier to desrcibe radiation using
the Stokes parameters $I$, $Q$, $U$, and $V$. Under conditions of QL
propagation, we get $Q=U=0$, and the $x$- and $o$-mode waves are
circularly  polarized in opposite senses. Thus we obtain $V=I_x-I_o$,
and the polarization  is circular having the sense of the dominant
mode. In the case of QT propagation, we obtain $Q=I_x-I_o$ and
$U=V=0$, and the polarization is linear.  However, because the Faraday
rotation in the corona is large, we can detect  linear polarization
only if we use receivers of much narrower bandwidth than  those
currently available. There is only one observation of linear
polarization at microwaves from active regions; it was accomplished by
\cite{AlD94} who used a multichannel spectral line receiver of very small 
bandwidth.

Under the approximation of geometrical optics (not very low values for
the magnetic field and electron density) when conditions change along
the radiation path, the polarization of the $x$- and $o$-mode waves
will change accordingly. Therefore when a transverse field region is
crossed, the sense of polarization changes because of the change of
the sign of the  longitudinal magnetic field. This is valid when the
coupling between the $x$- and $o$-mode waves is weak (i.e. they
propagate independently).  However, as both the coronal  magnetic
field and density decrease with height, the differences between the
characteristics of the $x$- and $o$-mode waves decrease, and hence
their mutual coupling increases. In the strong coupling regime, the
waves are not independent and the polarization does not change along
the path but attains a limiting value, even if a transverse field
region is crossed \citep[e.g.][]{Cohen60,Zheleznyakov70}. Therefore,
data of circular polarization do not necessarily reflect the polarity
of the magnatic field at the source of radiation
\citep[e.g.][]{Alissandrakis84,Alissandrakis93b,Shain17}.

The refractive index of the unmagnetized plasma is
$n=[1-(\nu_{p}/\nu)^2]^{1/2}$ where $\nu_{p}$ is the electron plasma
frequency. At low frequencies it can become much smaller than unity
which could trigger refraction and total reflection effects. Total
reflection of the radio waves will occur when $\nu=\nu_{p}$.
Refraction modifies the ray paths and also decreases the brightness
because the rays  move away from regions of high density and the
optical depth  becomes less than unity \citep{Alissandrakis94}. Generally, 
refraction is not important unless the optical depth between the region 
of total reflection and the observer is small. Its effect becomes more 
serious when large-scale density inhomogeneities are present in the corona 
and inner heliosphere (e.g.  coronal holes, streamers, slow or fast solar wind
streams). This can result in distorions and/or apparent position
offsets of the radio sources \citep[e.g.][]{Duncan79,Lecacheux89}. Ionospheric 
refraction can also significantly modify the apparent position of radio 
sources, sometimes more than several minutes of arc in the metric range
\citep[e.g.][]{Mercier96}.

When small-scale inhomogeneities are present between the observer and
the radio source, several scattering phenomena may take place. These
include spectral and angular broadenings that cause
frequency-dependent blurring in radio structures
\citep[e.g.][]{Bastian94} and decrease of the detected brightness temperature
at low frequencies \citep{Thejappa08}. Furthermore, anisotropic
scattering displaces radio sources \citep{Kontar19}.

\section{Free-free emission}

\subsection{Emissivity and absorption coefficient}

\subsubsection{Electron-ion free-free mechanism}

From the middle chromosphere upward, the free-free emission (or
bremsstrahlung) exclusively originates from electrons that are
diverted in the Coulomb field of ambient ions because they
are accelerated by the Coulomb force. The term ``free-free'' is due to
the state of the electrons; they are free both before and after the
interaction.

In the classical limit, the radiation of free accelerated charged particles
is described by Larmor's formula:

\begin{equation}
\frac{dP}{d\Omega} = \frac{q^2 a^2}{4 \pi c^2} \sin^2 \theta
\end{equation}
where $P$ is the power emitted within the solid angle $d\Omega$  by a
particle of charge $q$, mass $m$, and acceleration $a$ in the
direction $\theta$ relative to the acceleration vector. The total
radiated power is obtained after integration over solid angle:

\begin{equation}
P = \frac{2 q^2 a^2}{3 c^2} 
\end{equation}
Since $a \propto 1/m$, the power is $\propto 1/m^2$ and the proton
radiation can be ignored because it is much smaller than that of
electrons.  This conclusion holds for all radio emission
processes. Interaction between identical charges also does not produce
much radiation because radiation power is proportional to the
second derivative of the dipole moment of the system of charged particles,
which does not change when two identical particles interact. Consequently, 
only electron-ion collisions are relevant, and significant radiation is 
produced by the electrons only.

In a binary encounter between an electron of speed $v$ and an ion of
charge $Z$, the electron deviates from  its straight line path by an
angle $\theta$, which depends on its speed and the distance of the
encounter, called impact  parameter, $b$. In the corona there is a
large number of particles inside the Debye sphere and hence the ratio
of small-to-large angle encounters is  $\approx \lambda_D/r_c$, where
$\lambda_D$ is the Debye length and $r_c$ is the impact parameter that
produces a 90\degr\ deflection
\citep[e.g. see][]{Raulin05}. Consequently, small-angle collisions
dominate and the path of an incoming electron is determined primarily
by multiple interactions that yield small deflections, and therefore
low energy (radio, that is) photons are produced. In large-angle
encounters, high energy electrons may undergo large deflections that
yield the emission of high energy (X-ray, that is)
photons. Large-angle encounters become more important as we move to
cooler and denser deeper  layers of the solar atmosphere, because the
size of the Debye sphere decreases.

The calculation of the emission from free-free interactions is given
in detail by \cite{Rybicki79}; here, we only outline the
procedure. Since we are interested in the radio emission, the small-angle
approximation is appropriate for which the deflection of electrons can be
neglected, and therefore we assume the motion takes place along a straight
line where the electron and ion are separated by  $r = \sqrt{b^2 + v^2
t^2}$. We further use the dipole approximation to obtain the net acceleration
along the path, and therefore for a single collision an electron emits:

\begin{equation}
\frac{dW(b)}{d\omega} = \frac{8Z^2e^6}{3 \pi c^3 m_e^2 v^2 b^2 } \,\,\,\,\,\, \mbox{for} \,\,\,\, b \ll v / \omega
\end{equation}
(collisions at a given $b$ lead to emission only at $\omega < v/b$).
The total incoming flux of electrons with speed $v$ is $(n_e v)(2 \pi
b db)$, where $n_e$ is the electron number density. Then the free-free
emissivity (i.e.  emission per unit time, volume, and frequency) is:

\begin{equation}
\frac{dW}{dt dV d\omega} = n_e n_i 2 \pi v \int_{b_{min}}^{b_{max}} 
\frac{dW(b)}{d\omega} b db
\end{equation}
where $n_i$ is the ion number density. The limits of integration are determined
by $b_{min} = 4Ze^2/(\pi m_e v^2)$ corresponding to a 90\degr\ deflection,
and $b_{max} = v/\omega$, above which the emitted power is negligible.
When we combine eq. 18 and 19, we obtain:

\begin{equation}
\frac{dW}{dt dV d\omega} = \frac{16e^6 n_e n_i Z^2}{3c^3 m_e^2 v} 
\int_{b_{min}}^{b_{max}} \frac{db}{b} = \frac{16e^6 n_e n_i Z^2}{3c^3 m_e^2 v} 
\ln \left( \frac{b_{max}}{b_{min}} \right)
\end{equation}
Usually the above equation is written as

\begin{equation}
\frac{dW}{dt dV d\omega} = \frac{16 \pi e^6}{3^{3/2} c^3 m_e^2 v} n_e n_i Z^2
G_{ff}(v,\omega)
\end{equation}
where $G_{ff}(v, \omega)$ is the Gaunt factor \citep{Karzas61}.

The next step is to integrate eq. 21 over the velocity distribution
of the electrons. In radio astronomy, a thermal distribution is used
in most cases, and the calculation yields

\begin{equation}
\eta_{\nu} = \frac{2^5 \pi e^6}{3 m_e c^3} \left( \frac{2\pi}{3m_ek_BT_e} \right)^{1/2} n_e n_i Z^2 G_{ff}(T_e, \nu)
\end{equation}  
where the emissivity is now denoted by $\eta_{\nu}$, and $T_e$ is the 
electron temperature. The emissivity is proportional to the product
of the electron number density with the ion number density. The decrease
of emissivity with increasing temperature comes from the decrease of
$dW(b)/d\omega$ with increasing relative speed $v$ of the electron-ion
pairs (eq. 18).

Using the emissivity and the Rayleigh-Jeans law, $B_{\nu}(T_e) = 2k_BT_e\nu^2/c^2$,
we obtain the absorption coefficient:

\begin{equation}
k_{\nu} = \frac{1}{3c} \sqrt{\frac{2\pi}{3}} \left(\frac{\nu_{p}}{\nu}\right)^2 \frac{4\pi e^4 n_e n_i Z^2}
{m_e^{1/2} (k_B T)^{3/2}} G_{ff}(T_e, \nu)
\end{equation}
which is also written as

\begin{equation}
k_{\nu} = \frac{9.78 \times 10^{-3} n_e n_i Z^2}{\nu^2 T_e^{3/2}}
\times \left\{ \begin{array}{ll}
              18.2 + \ln T_e^{3/2} - \ln \nu & (T_e < 2 \times 10^5 K) \\
              24.5 + \ln T_e - \ln \nu & (T_e > 2 \times 10^5 K) 
\end{array} \right. 
\end{equation}
where two expressions for the Gaunt factor have been used for conditions 
relevant to the solar atmosphere \citep{Dulk85}.

\subsubsection{H$^-$ free-free mechanism}

Free-free absorption can result not only from interactions between
ions and free electrons but also from free-free transitions of electrons in the
field of hydrogen atoms. The latter mechanism is often referred to as H$^-$ 
absorption. \cite{Stallcop74} has provided analytical expressions for the
H$^-$ absorption coefficient, $k_{H^-}$, from which we obtain:

\begin{equation}
k_{H^-} = 1.2754 \times 10^{-11} \frac{n_en_H \sqrt{T}}{\nu^2}e^{-\zeta(2.065K)}
\end{equation}
where $n_H$ is the hydrogen density and

\begin{equation}
K = 2.517 \times 10^{-3} \sqrt{T}
\end{equation}
and
\begin{equation}
\zeta(2.065K) = 4.862K(1-0.2096K+0.0170K^2-0.00968K^3)
\end{equation}
The contribution of H$^-$ absorption becomes non-negligible at wavelengths 
shorter than about 1 mm, where radiation is formed in cooler layers. 
\cite{Alissandrakis17} have estimated that at around electron temperature
of 6000 K, the H$^-$ opacity is about 10\% of the total opacity
\citep[see also][]{Wedemeyer16}.

\subsection{Polarization}

With the above treatment we did not take into account the magnetic
field and we also assumed that the index of refraction $n$ is
unity. If we relax these assumptions, we obtain equations which are
more complicated than eq. 22 and 24. However, simplifications are
often used \citep[e.g.][]{Kundu65}, which yield  a simpler expression for
the absorption coefficient:

\begin{equation}
k_{\nu} = \xi \frac{n_e^2}{n \nu^2 T_e^{3/2}} \frac{1}{(\nu \pm \nu_{ce} 
|\cos \theta|)^2}
\end{equation}
where $\xi$ is a slowly varying function of $T_e$ and $n_e$, $\nu_{ce}$ is 
the electron gyrofrequency, and $\theta$ the angle
between the magnetic field and the line of sight. The plus sign in
eq. 28 corresponds to the $o$-mode emission while the minus sign
corresponds to the $x$-mode wave. Therefore the $x$- and o-mode opacities are
slightly larger and smaller, respectively, than that of the unmagnetized
situation. The radiation forms in regions where the temperature increases
with height, and so does the brightness temperature. Consequently, we expect
weak polarization in the sense of the $x$-mode.

For uniform thermal material and using eq. 28, we derive that the degree of 
circular polarization $\rho_c$ of the optically thin free-free emission is:

\begin{equation}
\rho_c = \frac{V}{I} = \frac{2 \nu_{ce} \cos \theta}{\nu}
\end{equation}
where the longitudinal component of the magnetic field, $B \cos \theta$, is
involved because $\nu_{ce} = eB/(2 \pi m_ec)$. However, one should take 
into a ccount that the density and
magnetic field are not constant along the line of sight, and that
the emission should not necessarily be optically thin. However,
in the weak field limit \cite{Grebinskij00} and \cite{Gelfreikh04} have
shown that one may nevertheless constrain the coronal magnetic field from
spectrally resolved data of free-free radiation.

\subsection{Observations of free-free emission}  

At most frequencies and locations (exceptions include regions with
strong coronal magnetic fields observed at microwaves, see
\S5), the non-flaring Sun produces radio emission via the free-free mechanism.
At low frequencies the corresponding emission is optically thick in the
corona (though refraction may bring $\tau_{\nu}$ below unity at metric 
wavelengths), while at higher frequencies there is a mixture of optically
thick radiation from cool chromospheric plasmas together with contributions
from hot coronal plasmas (can be either optically thin or thick) in active 
regions.

\subsubsection{Spectrum}

A model of the spectrum of the microwave brightness temperature of free-free
emission is presented in the left column of fig. 1 \citep[from][]{Gary94}. The
calculation was performed for coronal conditions ($T_e \approx 1.5 \times 10^6$
K, fully ionized H and 10\% He) neglecting magnetic field and assuming that
the index of refraction is 1. Then by using eq. (28) and the definition of
optical depth, $\tau_{\nu}$, we obtain

\begin{equation}
\tau_{\nu} \approx 0.2 \frac{\int n_e^2 dl}{\nu^2 T_e^{3/2}}
\end{equation}

The spectrum is flat where the emission is optically thick and the
corresponding $T_b$ is merely the electron temperature of the corona (see eq. 
15). At high frequencies, the coronal radiation becomes optically thin and
the brightness temperature decreases as $T_b \propto \nu^{-2}$ (see eq. 14 and 
30). The brightness temperature spectrum we show in the right
panel of fig. 1 comes from observations obtained with the Owens Valley
Solar Array (OVSA) at 16 frequencies in the range 1.4-8 GHz \citep{Gary87} 
and is consistent with the above interpretation. More recent observations
by \cite{StHilaire12} have confirmed the above results.

Recently \cite{Rodger18} and \cite{Rodger19} have shown that the spectral
gradient of millimeter free-free emission can be used for the diagnosis
of the optical depth of either isothermal or multithermal material provided
a correction is introduced to compensate for the change of the Gaunt factor
over the observed frequency range.

There is a long tradition of comparisons of brightness temperature
spectra of  the free-free emission that span a wide range of
frequencies (from sub-mm to  microwaves) with either standard
one-dimensional atmospheric models or  with the computed radio
brightness resulted from EUV observations. The reader is  referred to
the review by \cite{Shibasaki11}  and to the paper about the
solar atmosphere written by Alissandrakis in this issue for a
detailed discussion of the subject.

\subsubsection{Imaging observations of the non-flaring Sun}

The quiet Sun radio emission comes from the free-free mechanism. At frequencies
$\gtrsim 20$ GHz the corona is optically thin everywhere and radio emission 
probes the chromosphere. The first high-resolution images of the quiet Sun 
in the millimeter range were obtained by \cite{White06} and 
\cite{Loukitcheva06} who used the Berkeley-Illinois-Maryland Association Array
(BIMA) to obtain $\sim 10\arcsec$ resolution. With the advent of ALMA
a new generation of high-resolution millimeter-wavelength images has been
forming \citep[e.g.][]{Shimojo17a,Shimojo17b,Bastian17,Yokoyama18,Nindos18,Loukitcheva19,Jafarzadeh19,Molnar19,Patsourakos20,Wedemeyer20} and an example is 
presented in fig. 2. The figure indicates that the
chromospheric network, delineated in the AIA 1600 \AA\ image, is the
dominant structure in the radio images.  The AIA images of fig. 2 have
been degraded to the resolution of the ALMA images hence the size of the
network is similar in all three wavelengths. The chromospheric network is also 
visible at microwaves \citep{Kundu79,Gary88,Gary90}. Subsequent observations  
\citep[e.g.][]{Bastian96,Benz97,Krucker97} confirmed that result. 

In active regions, free-free emission is produced by the plage and
coronal loops. At high microwave  frequencies ($\nu >$ 3 GHz), the
active region free-free emission is optically  thin (for an example
see the 17 GHz image of fig. 3); at such high frequencies the only
regions of the non-flaring corona that are optically thick are
those with strong magnetic fields ($>$ 400 G) where gyroresonance
opacity is significant. \cite{White99} points out that large active regions
almost always contain optically thick regions at 1.5 GHz due to
free-free opacity, but their free-free opacity is never optically
thick at 5 GHz.

The observed brightness temperature of the free-free emission may fall
below the coronal electron temperature  not only at microwaves,
but also at metric wavelengths.  The corresponding fall of the
optical depth below unity at metric wavelengths is attributed to
scattering and refraction effects (see \S2.2). Note, however, that
the recent analysis of quiet Sun images obtained with the Low Frequency
Array (LOFAR) in the range 25-79 MHz indicates the presence of higher brightness
temperatures, of the order of 1 MK \citep{Vocks18}. 

Fig. 3 \citep[see also][]{Mercier09,Mercier12} shows images of the Sun at
150-432 MHz obtained with the  Nan\c{c}ay Radioheliograph (NRH). These
frequencies probe altitudes from the upper transition region
to the low corona. At the highest frequencies (327-432 MHz) the most
prominent feature is a coronal hole that appears as a depression south
of the disk center. Its contrast decreases at lower frequencies in
agreement with previous \citep[e.g.][]{Lantos87,Lantos99b} and more
recent observations \citep{Rahman19}. However, the dark-to-bright
transition at low frequencies cannot be easily reproduced in model
computations \citep{Rahman19}. Furthermore \cite{Mccauley19}
reported values up to 8\% for the polarization of coronal holes. 

Fig. 3 also shows that the similarity of the soft X-ray image with the
radio images decreases with frequency. This is not due to spatial resolution
differences only. Apart from radio refraction effects, the optical depth of 
the radio emission  is larger than that of the X-ray emission which is always 
optically thin. Consequently the radio images probe higher coronal layers
and lower-lying structures that emit soft X-rays are obscured by dense
overlying material. With similar arguments one can interpret the little
resemblance between the NRH images and the Nobeyama Radioheliograph (NoRH) 
image at 17 GHz.

\subsubsection{Imaging observations of flares and CMEs}

The hot soft-X-ray-emitting coronal material that fills flaring loops
is expected to produce optically thin free-free emission at millimeter
wavelengths and in the decay phase of microwave flares
\citep[e.g.][]{Kundu82,Hanaoka99}. Thermal free-free emission has also 
been detected in weak transient brightenings observed at microwaves 
\citep{White95} and millimeter wavelengths \citep{Nindos20}. For the sake of 
completeness we note that (i) nonthermal emissions have occassionally also 
been reported in the decay phase of flares \citep[e.g.][]{Kundu01c,Kundu04}, 
and (ii) sub-THz emission may originate from optically thick and 
relatively hot free-free sources located at the chromospheric footpoints of 
flaring loops \citep[][]{Morgachev18,Morgachev20,Kontar18}.

CME material is also expected to radiate optically thin free-free emission.
Ideally, one anticipates that the free-free-emitting structures will be
similar to those appearing in coronagraphs because both are associated with
multi-thermal plasmas and depend on the electron emission measure (radio 
frequencies) or column density (white-light data). But the free-free
emission of CMEs should be weak because of their low densities and high
temperatures (see eq. 30) and often obscured by stronger nonthermal emissions. 
The published reports on thermal free-free CME emissions at low
frequencies are rare. The reader is referred to the articles by 
\cite{Sheridan78} and \cite{Gopalswamy92} for early detections with the
extinct Culgoora Radioheliograph. \cite{Maia00} imaged CME fronts at
frequencies between 164 and 450 MHz. The radio source motions matched
those of white-light CME fronts and their brightness temperature 
($\sim 10^4$ K) implied thermal emission. However, their spectral 
characteristics and polarization were not consistent with such interpretation.
Thermal emission from CMEs has also been reported at 109 MHz by 
\cite{Kathiravan02} and \cite{Ramesh03} (Gauribidanur Radioheliograph 
observations). The thermal free-free CME emission, when detected, can be
used for the calculation of the CME mass. Such calculations provide results
similar to those obtained from white-light data \citep{Gopalswamy92}.

Prominences can be observed not only in H$\alpha$ but also in radio.
The H$\alpha$ emission strongly depends on the plasma temperature but
since prominence material is dense ($\sim10^{10}-10^{11}$ cm$^{-3}$)
and cool ($\sim 8000$ K), it produces optically thick thermal
free-free microwave  emission \citep[e.g.][]{Gopalswamy98}, which can
be easily detected beyond the disk. But quiescent filaments,
i.e. quiet promineces seen on the disk, are associated with brightness
depressions in the microwave range \citep[e.g.][]{Drago01} and
sometimes at decimeter-meter wavelengths
\citep[e.g.][]{Marque99,Marque04}. At microwaves, the angular
resolution is  not as good as in H$\alpha$ but, thanks to the
continuum nature of the  free-free emission, the microwave data have
the ability to probe the prominence material even at relatively high
temperatures which is not  always possible in H$\alpha$. From the
radio data one can calculate the prominence mass
\citep[e.g.][]{Gopalswamy98}, but such estimates ignore possible
downflows of plasma from the rising prominence. Several events
of prominence eruptions at microwaves (see fig. 4 for an example) have
been reported
\cite[e.g][]{Hanaoka94,Gopalswamy96a,Gopalswamy03,Hori00,Uralov02,GUZ06,Alissandrakis13,Fedotova18,Huang19}. In most cases the  eruptive prominence detected
in radio eventually evolves into the core of the white-light CME.

\section{Types of gyromagnetic emission}

Gyromagnetic emission is generated when free electrons are accelerated
or/and change their velocity direction
in a magnetic field due to the influence of the magnetic component of
the Lorentz force (in solar plasmas, there are no macroscopic electric
fields,  except probably  in current sheets). An electron with
velocity components $v_{\parallel}$ and $v_{\perp}$  parallel and
perpendicular to the magnetic field, respectively, will be accelerated
perpendicular to both $v_{\perp}$ and $B$. Its acceleration, $a$, is:

\begin{equation}
a = \omega_{ce} v_{\perp}
\end{equation}
where $\omega_{ce}$ is the electron gyrofrequency

\begin{equation}
\omega_{ce} = \frac{eB}{m_e c}
\end{equation}
For non-relativistic speeds, the total power emitted by the electron is
provided by the Larmor formula (eq. 17) which yields:

\begin{equation}
P = \frac{2e^2}{3c^2} \omega_{ce}^2 v_{\perp}^2
\end{equation}
This expression requires modification when the electron speed is not
small compared to the speed of light, $c$. Then the power of the electron
is given by the relativistic Larmor formula:

\begin{equation}
P = \frac{2e^2}{3c^2} \gamma^2 \omega_{ce}^2 v_{\perp}^2
\end{equation}
where $\gamma$ is the Lorentz factor.

It is straightforward to prove that the total emitted power, $P$, is
Lorentz  invariant. However, its angular distribution $dP/d\Omega$ is
not. In the electron rest frame, the radiated power per solid angle is
given by eq. 16 and the radiation pattern is the classical dipole
pattern which shows two lobes with power proportional to $\sin^2
\theta$. We assume that the relative velocity between the electron
rest frame and the observer rest frame is along the $x$-axis, and we
use spherical coordinates such that the angle $\theta$ is measured
with  respect to $x$-axis while the angle $\phi$ is the angle between
the acceleration and the projection of the line from the charge to the
observer  onto the plane that is perpendicular to the velocity. Then
for the emitted  power per solid angle in the rest frame of the
observer, the calculations  give:  

\begin{equation}
\frac{dP}{d\Omega} = \frac{e^2a^2}{4\pi c^2} \frac{1}{(1 - \beta \cos 
\theta)^4} \left[ 1 - \frac{\sin^2 \theta \cos^2 \phi}{\gamma^2(1 - \beta
\cos \theta)^2} \right]
\end{equation}
where $\beta=v/c$. There is a strong dependence on the $1-\beta
\cos \theta$ factor in the denominator which dominates when $\theta
\rightarrow 0$ and $\beta \rightarrow 1$. In other words, the observer
detects strong radiation in the forward direction with respect to the 
motion of the electron; this is called relativistic beaming. Therefore, 
in the relativistic case, we obtain strongly beamed emission along the 
direction of motion which, in turn, is perpendicular to the acceleration. 
The width of the beam where the emission is concentrated is $2/\gamma$. 
This means that the signal detected by the observer appears more and more 
sharply pulsed as the energy of the electron increases. Beaming plays 
important role in the observed spectrum emitted by a single electron.

The above discussion is valid for an electron radiating in vacuo.
In the presence of ambient plasma, we should take into account the
influence of the index of refraction $n$ on radiation. In that case,
the width of the emission beam is $\theta=2(1-n^2\beta^2)^{1/2}$. If
$n \approx 1$ we return to the vacuum case. But if $n \ll 1$,
then for the ultra-relativistic case ($\beta \sim 1$) we obtain
$\theta=2(1-n^2)^{1/2}=2\nu_p/\nu$ for a cold plasma. Therefore at low
frequencies, the medium quenches wave propagation and plasma effects
dominate beaming effects. The decrease in beaming takes place because
the electron cannot ``catch up'' to the wave it just emitted to reinforce
it with yet another emission wave. 

In the limit of non-relativistic speed (see left column of fig. 5),
the electron gyrates with frequency equal to the classical
gyrofrequency (eq. 32)  which is independent of the speed. An observer
will detect a sinusoidally  varying electric field which has a period
of $2 \pi / \omega_{ce}$. In that case, the power spectrum is a single
line (cyclotron line) at the   gyrofrequency. When the electron energy
increases, mild beaming is initiated and the observed temporal
variation of the electric field becomes  non-sinusoidal. In such
circumstances, when the electron energy corresponds to quiescent solar
coronal temperatures ($\sim10^6$ K; see middle column of fig. 5), the
power spectrum consists of lines at frequencies equal to small integer
multiples (called harmonics) of the gyrofrequency. In that case,
gyromagnetic  emission is called gyroresonance emission. For a mildly
relativistic  electron (energy of a few tens of keV to a few MeV)
there is even more  beaming, and there is power in a wide frequency
range at harmonics of the gyrofrequency from about 10 to 100. Now the
lines exhibit Doppler broadening and start blending together.  This
type of gyromagnetic  emission is called gyrosynchrotron
emission. Finally, when the electron is highly relativistic (right
column of fig. 5) the electric field's temporal variation becomes
highly non-sinusoidal and there is power at a large number of
harmonics, up to more than $s \sim \gamma^3$, which overlap to yield
continuum emission. The frequency $\nu=\gamma^3 \nu_{ce}$ that
corresponds to the maximum synchrotron emission of a single electron
(see fig. 5) is sometimes referred to as the  ``characteristic
frequency of synchrotron emission''.

The above discussion is valid for a single electron in the presence of
uniform magnetic field. However, in the corona, even along a single
line of  sight, the magnetic field is not uniform, but generally
decreases with height. The non-uniformity of the field
together with the spread of the distribution function of the electron
energy tend to smear out the spectral lines into an, essentially,
continuum emission.

Single expressions for the gyromagnetic emission and absorption
coefficients that are valid for all electron energies are not
available. Instead, expressions have been derived for separate
electron energy regimes.

\begin{itemize}

\item At low, non-relativistic energies ($\gamma -1 \ll 1$), the
electron velocity distribution in the corona is thermal and the 
resulting gyroresonance emission  is the primary emission mechanism above
sunspots with strong magnetic fields at microwaves.

\item In the case of gyrosynchrotron emission from mildly relativistic
electrons ($\gamma -1 \sim 1-5$), both thermal and non-thermal
electron  energy distributions have been used. Gyrosynchrotron
emission is the primary  incoherent emission mechanism in solar flares.

\item Synchrotron emission comes from ultra-relativistic
($\gamma-1 \gg 1$) electrons. It is well-known that synchrotron
emission is relevant in neutron stars, and some extra-galactic
sources. In the Sun, it is debated whether it may contribute to the
impulsive submillimeter component of some flares \citep{Trottet08}.

\end{itemize}

\section{Gyroresonance emission}

\subsection{Optical depth}

Gyroresonance opacity for electrons with a thermal
energy distribution has been discussed in several textbooks
\citep[e.g.][]{Zheleznyakov70,Melrose80}.  More recent detailed
reviews about the physical mechanism and the properties of
gyroresonance emission can be found in \cite{White97}  and
\cite{Lee07}. The gyroresonance absoption coefficient from
non-Maxwellian quasi-steady-state electron distributions has been
calculated by \cite{Fleishman14} but their predictions have not been
tested against  observations yet.

The absorption coefficient decreases quickly at frequencies not equal to
the resonance frequencies 

\begin{equation}
\nu = s \nu_{ce}.
\end{equation}
where $s$ is the harmonic number. In units of the scale height of the 
magnetic field, $L_B = B/|\nabla B|$, the frequency width of the resonances is 
$\sim v/c$, where $v$ is the speed of the emitting electron. In other words, 
when we observe at a fixed frequency, $\nu$, gyroresonance opacity becomes 
appreciable only at those points along the line of sight where $\nu_{ce}=
\nu/s$. Therefore, the magnetic field, electron density and temperature are 
almost constant across gyroresonance layers.

The exact expression for the optical depth $\tau$ of a gyroresonance layer
has been provided by \cite{Zheleznyakov70} and will not be repeated here.
The optical depth depends on many parameters but its most sensitive dependence
is on the angle, $\theta$, between the magnetic field and the line of sight.

In fig. 6 we show calculations \citep[taken from][]{White97} of the
optical  depth, at 5 GHz, of the second, third, and fourth gyroresonance 
coronal layers versus  the angle $\theta$. In the figure the
strong dependence of gyroresonance  optical depth on $\theta$ is
evident and shows better at small angles where the opacity drops
quickly in both modes. The second and third harmonics are optically
thick at most intermediate angles in both $x$- and $o$-mode. The
fourth  harmonic is optically thin in both polarizations. Harmonics
greater than the fourth have negligible opacity.  Only $o$-mode
emission is possible from the first harmonic, and this can happen if
the local cyclotron frequency exceeds the local plasma frequency. The
$x$-mode emission from the first harmonic does not propagate out into
the corona because the reflecting point of the $x$ wave is located
higher in the  corona (i.e. closer to the observer) than the $s=1$
layer. At a given  harmonic and angle, the $o$-mode opacity is about
an order of  magnitude smaller than $x$-mode opacity. For a given mode
and angle,  the transition from harmonic $s$ to harmonic $s+1$
decreases the opacity  by about two orders of magnitude.

\subsection{Structure of gyroresonance sources}

The structure of gyroresonance sources is determined to a large extent
by the number of harmonic layers that lie above the base of the
transition region (TR).  The magnetic field decreases with height and
therefore higher harmonic layers are located above lower ones. In the
case of magnetic field decreasing away from the center of a sunspot,
the height of a  given harmonic layer decreases with the distance from
the center.

The brightness temperature of a given harmonic layer depends on the
electron temperature at the height where it is located and on its
optical depth. Let us consider the case of a symmetric sunspot.  When
the harmonic layer is optically thick, the brightness temperature will
peak around the center of the spot. Away from the spot center, the electron
temperature   decreases, at first slowly and then fast as the harmonic
layer reaches regions with higher gradient of electron
temperature. Consequently, the source will have a flat top and sharp
borders. When the source is away from disk center the maximum
intensity is located toward the limb because the angle $\theta$
attains its highest values there. Furthermore, projection effects will
result in a faster drop of the brightness temperature in the direction
of the limb and a smoother drop toward disk center.

Due to the strong dependence of the gyroresonance opacity on the angle
$\theta$, a harmonic layer can become optically thick or thin at a
given frequency and heliocentric distance, depending on the position
within the source. The opacity is zero at $\theta=0$, thus there is
always a region around $\theta=0$ where the harmonic layer is
optically thin. This region can have considerable effects
on the structure of the source: it shows up as a region of small
intensity and it can result to a ring-shaped or a crescent-shaped
harmonic. When the spot is at disk center, this zero-$\theta$ region
is located at the center of the source, but as the source moves
toward the limb, the zero-$\theta$ region is displaced. In the
current-free case, where the field lines project radially on the
photosphere, this displacement is toward the disk center.  On the
other hand, when the field is not potential, the twist of the field
lines introduces an additional displacement in the direction
perpendicular to the direction of the center.  Thus a microwave source
associated with a nonpotential field will appear rotated with respect to
a source associated with a potential field.

Since the opacity is significantly greater in the $x$-mode than in the
$o$-mode, the same source may appear very different in the two modes,
and therefore such differences will be prominent in the circular
polarization maps. In general,  the source is essentially
unpolarized in regions where both modes are optically thick but
circular polarization  may go near the 100\% level if the
$x$-mode is optically thick and the $o$-mode is optically thin.

To illustrate the structure of microwave gyroresonance sources, we
will present model computations of both the gyroresonance and
free-free  emission resulted from a dipole magnetic field. The
magnetic moment of the dipole is $8 \times 10^{30}$ erg/G and is
located vertically below the photospheric disk center at a depth of $2
\times 10^4$ km. With these  parameters, the value of the maximum field
at the photosphere is 2000 G.

For the computation of the $x$- and $o$-mode emissions, we also need
to know  the electron temperature, $T_e$,  and density, $n_e$, in the
TR and the low corona. We have used the same approach
as in \cite{Alissandrakis80}: both the electron temperature and
density change only with height; the  temperature is determined by
constant conductive flux, $F_c$, and the density  by hydrostatic
equilibrium. The model is specified by an $F_c=2 \times 10^6$ erg
cm$^{-2}$ s$^{-1}$ and a density of 10$^{10}$ cm$^{-3}$ at $T_e=10^5$
K; the 10$^5$ K level is located at a height of 2000 km above the
chromosphere. Below 10$^5$ K and down to $2 \times 10^4$ K, the
temperature is determined by the model of Cheng and Moe (1977), while
above $2.6 \times 10^6$ K it is taken as constant. Models of
gyroresonance emission can be found in several other publications
\citep[e.g.][]{Zlotnik68a,Zlotnik68b,Gelfreikh79,Alissandrakis80,AlK84,Holman85,Kruger85,Hurford86,Brosius89,Lee93,Lee98,Lee99,Gopalswamy96b,Nindos96,Nindos02,Vourlidas97,Tun11,Kaltman12,Wang15,Nita18}.

Fig. 7 shows the $x$- and $o$-mode brightness temperature as a function
of distance from the center of the model sunspot that results from the
dipole field. Fig. 8 shows the resulted $I$ and $V$ emission.
The positions at which the harmonic layers cross the base 
of the TR are marked on the figures, so that the contribution
of each harmonic can be identified. At the highest frequency, 11.2 GHz, only
the fourth and third harmonic layers are above the base of the TR.
Moreover, only the third has some contribution and this is at the
$x$-mode only. Here we can clearly see the zero-$\theta$ region near the
center of the source.

The second harmonic shows up at 7.5 GHz, but it is optically
thick only in the $x$-mode; the third harmonic has still a small optical
depth and the source in the $x$-mode appears like a disk, surrounded by
a ring. Both the disk and the ring have sharp borders, a consequence of the
steep TR as noted above. The $o$-mode emission is very weak,
which results in a circular polarization of almost 100\%.

At 5 GHz the third harmonic becomes optically thick at the $x$-mode
and the source has the shape of a disk. On the contrary this as well
as the second harmonic show extended zero $\theta$ regions in the
$o$-mode emission, which appears as a bright ring surrounded by a
weaker one. The ring structure is also present in the total intensity
while the circular polarization has a disk-ring structure. Contrary to
the distribution at 7.5 GHz, the minimum in the  polarization does not
occur just outside of the second harmonic, but it is located at the
region of maximum brightness of the ordinary emission.

At frequencies lower than 5 GHz, there is contribution from the first
harmonic, as well. This has practically no effect on the $x$-mode (see
the relevant comments in \S5.1), but it serves to fill the gap of the
zero-$\theta$ region in the $o$-mode (the local cyclotron frequency
was larger than the local plasma frequency, therefore $o$-mode emission
from the first harmonic was possible). Thus the disk part of the
circular polarization becomes progressively lower as the frequency
decreases. The ring part is preserved still at 1.5 GHz due to the
contribution of the fourth harmonic in the $x$-mode. In general, the
circular polarization has a maximum around 5 GHz decreasing toward
higher frequencies due to the decrease of the brightness temperature
of both modes and toward lower frequencies due to the decrease of
opacity difference of the modes. The brightness temperature  in
total intensity increases as the frequency decreases due to the increase 
of both the opacity and the height of the harmonic layers.

\subsection{Observations of  gyroresonance emission}

The expected properties of gyroresonance sources discussed above, have
been   confirmed by both high spatial resolution observations at a few
frequencies \citep[e.g.][]{Kundu77,Alissandrakis80,Alissandrakis82,AlK84,Lang82,Kundu84,Gopalswamy96b,Nindos96,Nindos02,Vourlidas96,Vourlidas97,Lee98,Lee99,Zlotnik96,Brosius04,Brosius06,Tun11,Nita18}
as well as multi-frequency spectral observations \citep[e.g.][]{Akhmedov82,Alissandrakis93a,Alissandrakis19,Lee93,Gary94,Tun11,Kaltman12,Stupishin18}.

\subsubsection{Modeling of a well-observed sunspot source}

An example of a well-observed gyroresonance source
associated with a simple sunspot near disk center is provided in fig. 9
and 10.   Active region 4682 was observed with the RATAN-600 and the Very Large
Array (VLA). The RATAN-600 observations provided one-dimensional scans of the
Sun at several microwave frequencies, while the VLA provided
high-resolution maps at 5 and 1.5 GHz. The flux density spectra of the sunspot
source in $R$ and $L$ appear in fig. 9. In the same figure, we also
present model flux density spectra in $R$ and $L$ (the magnetic polarity of
the sunspot was negative and therefore the model $o$- and $x$-mode
emissions correspond to $R$ and $L$ polarizations, respectively).

The model we used was the same as the one described in \S5.2 with the
exceptions that (i) the magnetic field was computed through extrapolations
of the photospheric field provided by a vector magnetogram, and (ii)
pressure data in the TR from O IV lines were available for   
part of the sunspot region. The comparison of the observed and computed
flux density spectra allowed us to estimate the conductive flux, $F_c$ and the
height, $h_0$, of the base of the TR: we found $F_c=6 \times 10^6$
erg cm$^{-2}$ s$^{-1}$ and $h_0=2000$ km. 

Fig. 9 shows that at 11.1 GHz there is weak $o$-mode emission and
significant $x$-mode emission and the source is almost 100\% polarized.
Consequently, the third harmonic is located in the low TR. The
$x$-mode model fluxes increase from 15 to 7.5 GHz (note that
the RATAN 7.5 GHz $L$  data were not reliable) and reach a maximum
around 7.5 to 4.3 GHz, whereas the $o$-mode fluxes increase from
11.1 to 4.8 GHz and reach maximum around 3.7 GHz. At
frequencies lower than $\sim3.7$ GHz there is contribution to the
emission from the plage. However, both the $L$ and $R$ fluxes decrease
because the second and third harmonics have entered the upper
part of the TR where the temperature gradient is not large and cannot
compensate the effect of the $\nu^2$ factor which is involved in 
the computation of the spatially integrated flux density spectra.

The above discussion indicates that the third harmonic enters the
TR at $\nu \geq 11.1$ GHz and reaches the upper part of
the TR at about 7.5 GHz. The corresponding frequencies
for the second harmonic  are about 9.4 and 4.8 GHz. The combination of
these results yields a lower limit of 1400 G and an upper limit of 1800 G
for the magnetic field strength at the base of the TR. In the
upper TR and low corona the field is $\sim900$ G.

In fig. 10 (top four panels) we present the 5 GHz VLA maps in $I$,
$V$, $R$, and $L$.  The $I$ and $R$ maps and to some extent the $L$
map feature a crescent-shaped source which was rather asymmetric with
larger brightness temperatures  toward the south. There is also a
region of weak intensity in the $I$ and $R$ maps, which is attributed
to the zero-$\theta$ region; it does not appear exactly at the sunspot
center but is displaced northward. The maxima of the $I$, $R$, and $L$
maps show some clockwise rotation with respect to the limb
direction. These features can be explained in the framework of
nonpotential magnetic fields (see the discussion in \S5.2). The $V$
map indicates that there is little circular polarization where the
total intensity is high. It also shows three distinct maxima: one is
associated with the ``zero-$\theta$'' region and the others occur west
and east of the $I$ maximum, in locations where only the third
harmonic was in the corona.

In the bottom four panels of fig. 10 we present the best-fit models to
the VLA observations. The models were calculated using the $F_c$ and
$h_0$ values deduced from the spectral modeling and allowing the
force-free parameter, $\alpha$, used  for the magnetic field
extrapolations to take different values for the $x$-mode and $o$-mode
computations. This was consistent with the analysis of the vector
magnetograms which revealed that $\alpha$ was not constant over the
active region.  The effect of pressure variations into the computed
models was not important at 5 GHz, and this confirms that the magnetic
field is the dominant contributor to the emission. A comparison of the
observations and models of fig. 10 shows that the models reproduce key
features of the microwave morphology: the zero-$\theta$ and crescent
shape of the $R$ and $I$ maps, the clockwise rotation of the maximum
intensity  with respect to the limb derection, and the three local
maxima of the $V$ map.
  
\subsubsection{Gyroresonance versus free-free emission}

In active regions, at microwaves, gyroresonance opacity is competing
with free-free opacity. Free-free emission is ubiquitous in the corona
(see \S3.3) but  whenever sunspot-associated microwave sources of
coronal brightness temperature or/and high  degree of circular
polarization appear, they can be safely attributed to gyroresonance
emission. In these cases, the sunspot's photospheric magnetic field
should be strong enough to bring harmonic layers  of the gyrofrequency
above the base of the TR.  On the other hand, microwave free-free
emission is spatially extended, its brightness temperature is smaller
(because it is optically thin) and its degree of polarization is
small.  This situation is illustrated in fig. 11 where the spatial
scale of the 1.5 GHz   emission is consistent with the spatial scale
of the soft X-ray loops and the plage and comes primarily from
free-free emission.  Gyroresonance emission may have  some
contribution to the two bright sources of the 1.5 GHz image,  but
the rest of the 1.5 GHz emission (including the band of lower
emission which is more or less orthogonal to the soft X-ray loops
near the loop tops) should come  exclusively from free-free
emission. On the other hand, the 4.5 GHz image is radically
different: it shows localized bright  emission above the sunspot
due to gyroresonance in the strong magnetic  fields there. Note,
however, that some weak free-free emission can be  traced east of the
sunspot source even at 4.5 GHz. Multi-frequency observations of active
regions allowed  \cite{Gary87}  to observe the change in emission
mechanism from gyroresonance to predominantly free-free at about 3
GHz.

The highest frequencies where gyroresonance emission has been detected
lie in the range of 15-17 GHz
\citep[e.g.][]{Akhmedov82,White91,Alissandrakis93a,Shibasaki94,NindosKW00,Vourlidas06}.
At even higher frequencies, the results are not conclusive because
only a few imaging observations have been  reported. We note that an
active-region 34-GHz emission has been modeled as purely free-free by
\cite{Selhorst08}. At decimetric and metric wavelengths, we cannot
trace any sunspot-associated  sources because the free-free opacity is
so high that all the emission comes from regions well above sunspots.

\subsubsection{Gyroresonance as a tool to study coronal magnetic fields}

From eq. 36 we get that the magnetic field (in G) can be written as a function
of the harmonic number and frequency of observations through

\begin{equation}
B=357 \frac{1}{s} \frac{\nu}{1 \,\, \mbox{GHz}}
\end{equation}
and once we identify the harmonics which produce the emission, it is
easy to constrain the field strength in the TR and low corona. This
technique is especially powerful when multi-frequency data are
available and was  demonstrated in \S5.3.1 \citep[see also
e.g.][]{Akhmedov82,Alissandrakis93a,Lee93,Gary94,Korzhavin10,Tun11,Wang15,Nita18}. 
When only a single frequency is  available, we may identify the size of the
region, at the base of the TR, in which the field strength exceeds the
value that corresponds to the frequency of observations: at 15 GHz
\citep[e.g.][]{White91}) and 17 GHz
\citep[e.g.][]{Shibasaki94,NindosKW00,Vourlidas06} field strengths of
at least 1800 G and 2000 G, respectively, have been measured. Details
on the subject are given in this issue in the papers about coronal
magnetic field measurements by Alissandrakis and Bastian.

\section{Gyrosynchrotron emission}

\subsection{General remarks}

Gyrosynchrotron emission may arise in quite diverse solar
environments: 

(1) Solar flares. Gyrosynchrotron radiation from electrons that gyrate
in the magnetic field with energies  of tens to hundreds of keV is the
basic emission mechanism at microwaves. The literature is vast and selected
references will be provided in \S6.2-6.4, primarily for publications that
link modeling of gyrosynchrotron emission with observations. Gyrosynchrotron 
radiation can also be detected at millimeter wavelengths and is produced by 
electrons with energies of more than 1 MeV \citep[e.g.][]{White92,Kundu94,
Silva96,Raulin99,Silva16,Tsap18}.

(2) Weak transient brightenings, when observed at microwaves, may
sometimes show emission consistent with the properties of
gyrosynchrotron radiation \citep[e.g.][]{Gary97,Krucker97,Nindos99,Kundu06}.

(3) Gyrosynchrotron emission has been detected in a small number of
CMEs at decimetric and metric wavelengths \citep{Bastian01,Maia07,Tun13,Bain14,Carley17,Mondal19}.

In what follows we will put emphasis on the microwave gyrosynchrotron 
emission from flares because it is a mature topic that has attracted most of
the attention  on the subject, and because it demonstrates nicely the
properties of the gyrosynchrotron mechanism. More on gyrosynchrotron emission
from CMEs can be found in this issue in the paper about radio CMEs by 
Vourlidas.
  
Both the free-free \citep[e.g.][]{Bastian07} and gyroresonance 
\citep[e.g.][]{Preka88} emissions produced by ambient
thermal  electrons should be taken into account when discussing 
incoherent  emission of microwave bursts. Compared to gyrosynchrotron,
they both have negligible effects in the emission,  but they are
important because they may increase the optical depth in the
chromosphere and the low corona.

Gyrosynchrotron emission can be produced by electrons with either a
nonthermal or a thermal distribution; in the latter case
[\citep[e.g.][]{Gary89} and \citep[for more recent examples
    see][]{Fleishman15,Wang17}] they could be electrons heated due to
the flare. Usually the emission is first computed for a single
electron radiating in cold plasma \citep[e.g.][the ``parent'' of all modeling 
papers]{Ramaty69}, while for the thermal case,
\cite{Gershman60} considered small fluctuations in the thermal
equilibrium of a magnetized plasma described by the linearized Vlasov
equation. The resulting formulas for the emission and absorption
coefficients for an ensemble of electrons involve integration over the
distribution function and summation over harmonics. 

Simplified expressions have been provided by \cite{Petrosian81} and 
\cite{Dulk82}. They have a limited range of validity but they are
useful in some applications. The model of
\cite{Petrosian81} is valid at harmonic numbers below 10. But it
only deals with emissivity, which means that it can only  be applied
to high frequencies. \cite{Klein87} extended this model to the
absorption coefficient. The agreement with Ramaty's numerical
calculations was quite good, starting at low harmonic numbers. By
construction Klein's model completely smears out the lines, and it
is devised for an isotropic electron population. Furthermore, it does 
not provide handy formulas for analytical calculations. 

The model by \cite{Dulk82} is valid above the tenth harmonic of the
gyrofrequency (consequently if the magnetic field is 500 G it cannot
be used at frequencies lower than 15 GHz), for a spectral index
$\delta$ of an isotropic power-law distribution of radiating electrons
with  $2 \le \delta \le 7$, and for angles $\theta$ between the
magnetic field and  the line of sight with $\theta \ge 20\degr$. At
high harmonics  (above the 50th), the synchrotron approximation can be
used in cases where the  effects of high energy cut-off can be
neglected.

The flux spectrum is divided into an optically thick part (flux rises 
with frequency) and an optically thin part (flux falls with frequency).
Spectral maximum corresponds to the frequency defined by $\tau_{\nu} \sim 1$,
and usually occurs at low harmonics of the gyrofrequency. The optically thin
component of the spectrum is mostly shaped by the energy distribution of the
electrons. In the synchrotron approximation the spectral index $\alpha$ in the 
high-frequency part of the spectrum is

\begin{equation}
\alpha = \frac{\delta - 1}{2}
\end{equation}  
For mildly relativistic electrons, the approximation given by \cite{Dulk82} 
is often used:

\begin{equation}
\alpha = 0.90 \delta - 1.22
\end{equation}
A comparison of equations (38) and (39) indicates that the emission decrease
with frequency is steeper at mildly than at ultra relativistic energies.
This is because a highly relativistic electron radiates over a broader
frequency range than a mildly relativistic electron.

The optically thick part of the spectrum is influenced primarily by
the  effects of the ambient plasma and radiative transfer. For large
ambient  densities the refractive index reaches zero at low
frequencies (see the  discussion in section 4) and both the absorption
and emission coefficients  are suppressed. This leads to intensity
suppression at low frequencies and the shift of the
spectral maximum toward higher frequencies.  This effect is known as
the Razin effect
\citep{Razin60,Klein87,Belkora97,Melnikov08,Song16,Fleishman18}.
Furthermore, if the optical depth of the emitting electrons is larger
than unity, as it often happens at low frequencies, the intensity
spectrum falls below the emission coefficient spectrum. This is the
self-absorption effect  which makes gyrosynchrotron radiation to fall
off steeply with decreasing frequency. 

The $x$-mode is associated with higher emission coefficient than the
$o$-mode while the inverse holds for the source function. Therefore, the sense 
of polarization corresponds to $x$-mode in an optically thin region and to 
$o$-mode in an optically thick region. The degree of polarization increases 
with the angle $\theta$ between the magnetic field and the line of sight.

For the above discussion we assumed that the pitch-angle (i.e. angle
between the velocity of the electron and the magnetic field)
distribution of the radiating electrons was isotropic. This can be
achieved by  collisions or by wave-particle interactions. But in a
flaring loop there is little emission from electrons with small pitch
angles, so the emissions  produced by different electrons, some with
large and some with small pitch  angles, can be significantly
different. \cite{Fleishman03a} and \cite{Fleishman03b} have discussed
how  pitch-angle  anisotropies affect gyrosynchrotron emission. It is
known \citep[see][and references therein]{Fleishman98} that when
anisotropic pitch-angle distributions prevail, the absorption
coefficient can become negative and coherent electron cyclotron maser
emission is produced. In flares such emission has a typical timescale
of the order of tens of milliseconds which is much shorter than that
of gyrosynchrotron emission (order of tens of seconds). Therefore
these two types of emission can be distinguished from their different
duration. Moreover, there are cases that although the  pitch-angle
anisotropy significantly reduces the absorption coefficient, the
latter remains positive \citep{Fleishman03a}

\cite{Fleishman03a} and \cite{Fleishman03b} showed  that the changes
to the gyrosynchrotron spectrum due to  pitch-angle anisotropy are
larger for small values of the angle, $\theta$, between the magnetic
field and the line of sight (compare the two top panels of
fig. 12). The degree of polarization increases as the anisotropy of the
pitch-angle distribution becomes larger and may approach the 100\%
level in the optically thin limit (see middle row of fig. 12). A
similar trend is registered for the spectral index of the optically
thin part of the spectrum (see bottom row of fig. 12) when the 
angle $\theta$ is small.

\subsection{Gyrosynchrotron emission from model flaring loops}

We will present calculations of the gyrosynchrotron emission in a model
flaring  loop to illustrate the properties of gyrosynchrotron
radiation.  The models have been published by \cite{Kuznetsov11} and are 
based on the codes developed by \cite{Fleishman10}. Similar models 
have been developed by \cite{Simoes10} and \cite{Osborne19}.
The magnetic field of the model loop is produced by a
dipole  below the solar surface. The loop is located at the solar
equator and its orientation is characterized by a heliographic
longitude of 20\degr\ and an angle of 60\degr\ between the
magnetic dipole and the equatorial plane. The height of the loop is
10\arcsec, its radius at the top is  2\arcsec, and the footpoint
separation is 11.5\arcsec. The magnetic field strength at the
footpoints and the top of the loop is 800 G and 75 G,
respectively. The loop is filled with uniform ambient thermal plasma
with a density of 10$^{10}$ cm$^{-3}$ and a temperature of $2 \times
10^7$ K.

The energetic electrons have a power-law index of $\delta=4$ and low- and
high-energy cutoffs of 100 keV and 10 MeV, respectively.  The pitch-angle
distribution can be either isotropic or a  loss cone modeled by

\begin{equation}
g_{\mu}   \sim 
\left\{ 
\begin{array}{ll} 1, \,\,\,\,\,\,\,\,\,\,\,\,\,\,\,\,\,\,\,\,\,\,\,\,\,\,\,\,\,\,\,\,\,\,\,\,\,\,\,\,\,\,\, \mbox{for} \,\, |\mu| < \mu_c \\ 
\mbox{exp} \left[ -\frac{(|\mu|-\mu_c)^2}{\Delta \mu^2} \right],  \,\, \mbox{for}     \,\, |\mu| \geq \mu_c          
\end{array} \right. 
\end{equation}
where $\mu_c=\cos \alpha_c$, $\alpha_c$ is the boundary of the loss cone, and 
$\Delta \mu$ controls how sharp this boundary is. In the
models of fig. 13, $\Delta \mu =0.2$. The spatial distribution of energetic
electrons along the loop is given by

\begin{equation}
n_e \sim \mbox{exp}[-\epsilon^2(\phi-\pi/2)^2]
\end{equation}
where $n_e$ is the number density, $\phi$ is the magnetic latitude, and
$\epsilon$  is a dimensionless parameter controlling the degree of 
spatial inhomogeneity of $n_e$ along the loop 
(for $\epsilon=0$ the distribution is homogeneous). 

The gyrosynchrotron emission from the model loop is shown in 
fig. 13 for the cases of (i) energetic electrons with isotropic pitch-angle
distribution and constant number density, $n_e=3 \times 10^6$ 
cm$^{-3}$, along the  loop, (ii) same as (i) but with a
loss-cone pitch-angle distribution, and (iii) same as (ii) but with a
spatially inhomogeneous energetic electron  distribution  which is
controlled by $\epsilon=4$ and by a loop-top number density of $2.8
\times 10^8$ cm$^{-3}$. These parameters yield a footpoint number
density equal to the number density used in cases (i) and (ii).
 
Let us have a closer look at the models with homogeneous and isotropic
electron distribution (see row (a) of fig. 13). Similar models have
been  published by \cite{Preka92}, \cite{Bastian98}, \cite{Nindos00},
\cite{KunduNG04}, \cite{Simoes06}, and \cite{Costa13}. At 3.75 GHz,
the source is optically thick and traces out the spatial extent of
magnetic volume accessible to energetic electrons. At 9.4-34 GHz, the
emission is optically thin and shows compact sources associated with
the footpoints of the loop.

This picture agrees well with the properties of gyrosynchrotron
emission. The magnetic field is larger near the footpoints and
smaller at the loop top. At a fixed frequency, the harmonic number
varies from lower values at the footpoints to higher values at the loop
top.  The mean energy of the emitting electrons is proportional to the
effective temperature, $T_{eff}$ which in turn is 
\citep[according to the simplified expressions by][]{Dulk82} proportional to
$s^{0.5+0.085\delta}$. Therefore, higher energy electrons emit 
at the loop top, while lower energy electrons emit at the footpoints. In other 
words, the strong field near the footpoints favors the higher frequencies. 
Decreasing the observing frequency has approximately the same
effect for the gyrosynchrotron emission as increasing the magnetic
field. Consequently when the observing frequency decreases, we anticipate 
to obtain emission not only from the footpoints but also from a large
part of the flaring loop.

In the models with homogeneous and isotropic electron distributions,
changes to the model parameters have the following effects.

{\it Magnetic field strength}. A large magnetic field strength increases
the opacity and therefore decreases the electron number density
required to obtain the same optically thin flux. 
Furthermore it decreases the harmonic at which the electrons  
radiate at a given frequency. A larger field than that of the model is 
required to make the 9.4-34 GHz emissions optically thick and produce 
extended emission there.

{\it Loop thickness}. Changing the loop thickness increases the opacity
proportionally, and that effect increases the optically thin flux without
changing the optically thick flux significantly.

{\it Electron number density}. In the optically thin case, the optical
depth increases with the column density of the energetic
electrons. Changing the number density has little effect in the
optically thick case, but it affects the  frequency where the spectral
maximum occurs.

{\it Electron energy cutoffs}. A decrease of the upper limit to the
electron energies suppresses radiation at high frequencies which
requires very energetic electrons if the magnetic field  is not
large. An increase of the lower energy cutoff does not affect much the
high-frequency emission because the low energy electrons do not
radiate at high frequencies, but it increases the mean energy of the
electrons producing the 3.75 GHz optically thick emission and makes it
stronger.

{\it Viewing angle}. The viewing angle changes when we change the
orientation and location of the loop. In many cases, the changes
affect the microwave morphology significantly, because the
gyrosynchrotron mechanism depends strongly on the angle
between the line of sight and the magnetic field.

When there is a homogeneous density profile of energetic electrons
along the loop, the morphologies of the sources are similar in both
the isotropic and the loss-cone pitch-angle distribution cases
[compare  rows (a) and (b) of fig. 13]. However, there are differences
in intensity which show better in the spatially integrated spectra of
row (c) of the figure: at the optically thin frequencies, the emission
from the loss-cone distribution is lower than the emission from the isotropic
electrons by a factor of $\sim$ 2-6 [note also the differences in
maximum brightness temperatures between the maps, at a given
frequency, of rows (a) and (b)].  Furthermore, at 17 and 34 GHz, the
emission from the anisotropic population is more evenly distributed
along the loop than the emission from the isotropic population.  On
the other hand, at the optically thick frequencies the corresponding
intensities  are almost identical.

The interpretation of the above differences is as follows. The model loop
is located relatively close to disk center, where the angle $\theta$
between the magnetic field and the line of sight is small near the
footpoints, whereas near the loop top $\theta$ is large. When  
$\theta$ is small, the intensity at low frequencies is not sensitive 
to the pitch-angle anisotropy because the low-energy electrons contribute
most of the emission. The beaming effect (see \S4) for a single
low-energy electron is not large and therefore, the actual angular
distribution is not so important for the low-frequency emission.

The beaming effect becomes more prominent as the energy of the
emitting electrons increases, which results in the suppression of
higher frequency emission from the loss-cone distribution when
$\theta$ is small (we remind that close to the footpoints, the
electrons with loss-cone distribution are concentrated around a
pitch angle of 90\degr\ whereas the model field is almost parallel to the
line of sight).   When the distribution is anisotropic, this effect
results in the decrease of the footpoint emission which makes the
difference between the footpoint and looptop emission smaller than
that of the isotropic distribution.

When $\theta$ is large (i.e. near the loop top), the loss-cone
boundary, $\alpha_c$, falls to $\sim$ 20\degr\ and therefore the
loss-cone distribution does not differ much from the isotropic
one. Consequently, both distributions will produce very similar
emissions.

The emission from electrons with loss-cone pitch-angle distribution
and  inhomogeneous spatial profile of the electron density along
the loop [see row (d) of fig. 13] shows significant differences from
the cases discussed so far. The emission at 9.4, 17, and 34 GHz is
optically thin (the turnover frequency of the spectrum occurs at 7.6
GHz) but it peaks close to the loop top, and so does the emission at
3.75 GHz which is optically thick.  This change is attributed to the
larger concentration of energetic electrons at the loop top. Since the
relative contribution of the electrons close to the footpoints
decreases, the effect of the anisotropy discussed previously becomes
less prominent.

Modeling of gyrosynchrotron emission has gone a long way from the
pioneering publications by \cite{Klein84} and \cite{Alissandrakis84}.
In recent years it received  a major boost with the development of the
``GX$\_$Simulator'', an interactive IDL application which implements
the fitting scheme developed by \cite{Fleishman09} and the code by
\cite{Kuznetsov11} and allows the user to produce spatially-resolved
radio or X-ray spectra using realistic inputs for the magnetic field
and the  properties of both the energetic and ambient electrons
\citep{Nita15}. Results have appeared in several publications; some of
them have already been cited while references for others will be
provided below.

A final note about the determination of the magnetic field of flaring
loops is in place here.  The diagnostic strength of gyrosynchrotron,
albeit significant, is not as straightforward as that of
gyroresonance; for meaningful results one needs to combine
observations (ideally spectroscopic imaging ones) with detailed
modeling.    In spite of all the complications, modeling of individual
flares
\citep[e.g.][]{Nindos00,Kundu04,Tzatzakis08,Gary13,Gary18,Kuznetsov15,Fleishman16b,Fleishman16c,Fleishman18,Kuroda18}
showed that the magnetic field may lie from less than 200 G (loop top)
to about 1700 G (footpoints). Probably the most spectacular result was
obtained by \cite{Fleishman20} who modeled spectroscopic imaging
observations from Expanded OVSA (EOVSA) and found that the magnetic
field decayed at a rate of about 5 G s$^{-1}$ for two minutes.

\subsection{Observational examples}

Actual microwave observations of flares do not always show the simple
loop configuration used in the models of \S6.2. First of all, in some
cases  the spatial structure of the emission may not be resolved in
the radio maps. Furthermore, microwave sources may arise from pairs of
interacting loops of widely differing scales \citep{Hanaoka97,Nishio97,
  Grechnev06}. Configurations involving more complex loop systems have
also been revealed
\cite[e.g.][]{Kunduschmahl82,Kundu04}. Pre-eruptive flux rope
configurations have also been imaged at microwaves \citep{Wu16,Chen20} with
their emission coming, at least partly, from the gyrosynchrotron
mechanism.
 
The comparison between observations and models of the gyrosynchrotron
emission becomes possible in events with single-loop morphology. As an
example of such events, we consider  a microwave flaring loop
\citep{Nindos00}  observed by the VLA at 5 and  15 GHz (see fig. 14,
top panel). Additional spectral data  were obtained from the OVSA  at
several frequencies between 2 and 15 GHz; they revealed that the
turnover frequency was 5.4 GHz. At 15 GHz, the emission was optically
thin and was produced at the footpoints of the flaring loop, while the
5 GHz emission outlined the loop with most of it being optically thick
with a maximum close to the loop top.  In the middle and bottom panels
of fig. 14 we compare the observations with computations of
gyrosynchrotron emission from a model magnetic loop in order to
diagnose the conditions in the flaring loop. The best fit to the data
was reached with a model flaring loop with photospheric footpoint
magnetic field of 870 G. The thickness of the model loop was much
smaller than its footpoint separation. The energetic electrons were
characterized by an energy spectral index of 3.7, number density of
$7.9 \times 10^7$ cm$^{-3}$ as well as low- and high-energy cutoffs of
8 and 210 keV, respectively. In this model, the 5 GHz emission comes
from low harmonics of the gyrofrequency (3-7), while the lack of
electrons with energies higher than 210 keV was necessary to interpret
the absence of emission from the loop top at 15 GHz. That model [which
  is consistent  with the models presented in row (a) of fig. 13]
reproduced well both the  high-frequency part of the OVSA spectrum and
the  basic spatial structure of the VLA $I$ maps (propagation effects,
see \S2.2,  affected the structure of the $V$ maps, and therefore  its
comparison with  the model was not straightforward).

Observations show that in several cases the electrons that produce
gyrosynchrotron emission have often anisotropic and/or inhomogeneous
distributions. Such examples are the limb events presented by \cite{Kundu01a}. 
These flares were imaged by the NoRH at 17 and 34 GHz, and the emission
at both frequencies was extended and peaked close to the top of the
loop. On the other hand, spectral data from the Nobeyama Polarimeter
revealed that both the 17 and 34 GHz emissions were optically thin. 
A similar event was studied by \cite{White02}. \cite{Tzatzakis08} found that 
36\% of the events of an extended database of single-loop 
limb flares observed by the NoRH showed optically thin emission with
maximum close to the loop top. The morphology of these events is not
consistent with the morphology of optically thin sources from
homogeneous distributions of electrons.

\cite{Melnikov02} found that optically thin sources with loop-top
maxima can result from enhanced concentrations of accelerated electrons at
the loop top due to the transverse pitch-angle anisotropy of
the injected particles.  When the pitch-angle distribution of the injected
population is either beam-like (i.e. injection along magnetic
field lines) or isotropic, the resulted microwave emission peaks above
the footpoints.  \cite{Melnikov02} noted that another possible reason
for the concentration of energetic electrons near the loop top is the
enhanced losses of electrons close to the footpoints. This
possibility may occur either from Coulomb collisions if in the lower
part of the loop the plasma density is higher or from 
stronger turbulence there. \cite{Stepanov07} reported that strong
scattering of electrons by whistler waves can reproduce the
evolution of collimated streams of nonthermal electrons
observed by \cite{Yokoyama02}. \cite{Kuznetsov15} observed an
optically-thin loop-top source and argued that the strong
concentration of electrons near the loop top reflected the localized
particle injection  process accompanied by trapping and scattering.

Other publications that report pitch-angle anisotropies of the
electrons that emit gyrosynchrotron emission include \cite{Lee00},
\cite{LeeGS00},  \cite{Fleishman03c}; \cite{Fleishman06},
\cite{Altyntsev08,Altyntsev19},  \cite{Tzatzakis08},
\cite{Reznikova09}, and \cite{Charikov17}. Generally speaking, when
pitch-angle anisotropy is present it is not correct to derive the
energy spectrum of the energetic electrons from the slope of the
optically thin part of the gyrosynchrotron spectrum \citep[see][and
  the discussion in \S6.1]{Fleishman03a,Fleishman03b}. Instead, one
needs to resort to either forward fitting \citep[e.g.][]{Gary13} or 3D
modeling \citep[e.g.][]{Tzatzakis08,Nita15} in order to obtain
meaninful electron diagnostics from the radio emission.  

The study of the dynamics of flare microwave gyrosynchrotron emission
can provide important information about the kinematics of accelerated
electrons in flaring loops.  For example, analysis of the dynamics of
the spatial distribution of emission intensity, circular polarization
and frequency spectrum allows one to determine the localization of the
electron acceleration/injection region in a flare loop, as well as the
type of electron pitch-angle distribution in different parts of the
flare loop \citep[e.g.][]{Reznikova09,Melnikov12,Morgachev15}. Furthermore, 
the measured spectral dynamics of the microwave emission in the optically
thin part of the spectrum may provide important information on the
whistler  turbulence in the flare loop \citep[][]{Filatov17}.

Imaging spectroscopy can provide additional information to the study
of gyrosynchrotron emission and this is highlighted in fig. 15
\citep{Gary18} where imaging observations from the Expanded OVSA
(EOVSA)  of a partially occulted flare at 28 frequencies are
presented. Panel (a) of the figure shows the diversity of source
morphologies: as the frequency  increases the marginally resolved 
cusp-like source at frequencies between 7.4 and 8.9 
GHz gradually evolves toward a loop-like source while at frequencies
below 5 GHz two additional sources appear (see panel f), presumably
associated with the footpoints of a larger loop. The spectral modeling
presented in panels (b)-(e) yields the magnetic field and spectral
index of the electron energy distribution at the points marked in
panel (f).  This example shows that the combination of spatially
resolved radio spectra with realistic modeling can provide detailed
estimates of the dynamically evolving parameters in the flare
configuration. 

\subsection{Electron acceleration and transport}

In \S6.2 and \S6.3 we discussed gyrosynchrotron emission primarily
at a fixed time (presumably at the peak of the emission). However, flares 
are dynamic phenomena, and there is a large literature on the dynamics of
flare microwave emission with emphasis on the processes of electron 
acceleration and transport. Details are given elsewhere in this issue, 
thus this topic will be only touched on here.  

The study of the dynamics of flare microwave gyrosynchrotron emission can
provide important information about the kinematics of accelerated electrons
in flaring loops 

The combination of microwave and hard X-ray observations of flares with 
state-of-the-art 3D modeling provides a powerful diagnostic of accelerated
electrons \citep[e.g.][]{Fleishman16a,Kuroda18}. In the former study several
flares were analyzed which, instead of showing the usual broad-band 
gyrosynchrotron emission produced by electrons trapped in flaring loops, they 
showed narrow-band gyrosynchrotron spectra \citep[see also][]{Fleishman11,
Fleishman13}. The relationship of these bursts with hard X-rays together
with spectral modeling revealed that the trapped electron population was 
negligible and the radio emission originated directly from the acceleration 
sites which featured rather strong magnetic fields and densities. In the
\cite{Kuroda18} study the microwave and hard X-ray observations were
successfully fitted with a broken power-law spectrum that reproduced
the main characteristics of both emissions.

The most popular model for the study of electron transport during flares
is the ``direct precipitation/trap plus precipitation" (DP/TPP) model
\citep[see][and references therein]{Bastian98,Aschwanden02,Aschwanden04,
White11}. Energetic electrons with small pitch
angles travelling along appropriate magnetic field lines approach the
chromosphere where they are stopped by its dense and cool material.
Most of their energy heats the chromosphere, but a smaller fraction is
emitted in hard X-rays through the non-thermal thick-target free-free
mechanism. The coronal magnetic field traps electrons with large pitch angles  
inside the flaring loop where they emit gyrosynchrotron radiation.
However, eventually they will be scattered into the loss cone under the
influence of either Coulomb colisions or wave-particle interactions
and will precipitate into the chromosphere emitting additional hard X-ray 
radiation.

Using the TPP scenario one can study the effect of Coulomb collisions
on the energy of electrons \citep[see][and references therein]{Aschwanden04}.
The more energetic the electrons, the fewer collisions they undergo, and 
therefore the longer their lifetimes in the loop. In this way we can interpret
the frequency-dependent delays among microwave maxima, the usual lag of
microwave emission with respect to the hard X-ray emission, and the slower
decay of microwaves than hard X-rays.

Microwave emissions from either directly precipitating electrons 
\citep{Kundu01b,Lee02} or from electrons that have been efficiently scattered
\citep{Musset18} have also been detected. The microwave  emission from these 
populations does not have the same emissivity as the trapped electrons 
because their pitch-angle distributions are different. Hard X-rays do not come
exclusively from precipitated electrons; of course the thick-target
emission is more efficient, but trapped electrons also emit free-free
radiation, and this has been used to interpret long-duration hard
X-ray bursts \citep{Vilmer82,Bruggmann94}.  On the other hand,
microwave emission is also sensitive to the entire distribution of
electrons (both trapped and DP components), but the trapped component
will dominate the emission at a given frequency.

\section{Concluding remarks}

Incoherent solar radio emission is provided by the free-free,
gyroresonance, and gyrosynchrotron processes. Free-free radiation
dominates the quiet Sun and non-flaring active region emissions
with the exception of regions of strong fields above sunspsots where 
gyroresonance emission is large at microwaves. Gyrosynchrotron is the most 
important incoherent mechanism in flares.

Free-free opacity favors cool, dense plasmas, but if the density is
high enough then hot material can also produce bright radiation. Since
free-free  emission is ubiquitous in the Sun, it can be used to probe
the non-flaring solar atmosphere above temperature minimum. For this
task, the free-free emission has the advantage that it can be observed
from the ground and that it is not sensitive to processes affected by
ionization  equilibrium which characterize the EUV and X-ray
observations.

Gyroresonance opacity depends strongly on the magnetic field
strength and orientation. The emission is generated in thin layers above the
base of the TR that have practically constant magnetic field strength which is
determined by the condition that the observing frequency is equal to
low harmonics of the gyrofrequency. Coronal magnetic fields cannot be measured 
from the Zeeman effect; consequently multi-frequency microwave imaging 
observations of gyroresonance sources provide a unique tool for the 
determination of the three-dimensional structure of susnspot coronal magnetic 
field.

Unlike EUV and X-ray coronal emissions which are optically thin
everywhere, coronal radio emission can become optically thick due to
gyroresonance above regions of strong fields or due to free-free at
low frequencies. This means that radio data may allow us to probe
different layers in the solar atmosphere by  observing at different
frequencies. On the other hand, it is fair to say that radio images
cannot reach the spatial resolution and crispness of the images
obtained with some modern EUV instruments, for example the Atmospheric Imaging
Assembly aboard Solar Dynamics Observatory.

Gyrosynchrotron radiation is emitted at microwaves and millimeter
wavelengths from accelerated electrons of mildly relativistic energies
(i.e. from a  few tens of keV to a few MeV) as they move
in the coronal magnetic field. Gyrosynchrotron provides
powerful diagnostics of physical conditions  in flaring sources,
because it depends on the properties of both the magnetic field  and
the accelerated electrons, as well as the properties of the ambient
plasma. 

The diagnostic potential of the emission mechanisms discussed in this
paper has not been fully exploited yet. The basic reason is that until
relatively recently there was no instrument capable of performing
imaging spectroscopy over a wide frequency range. However, things have
been changing with the upgrade of both solar-dedicated (Owens Valley
Solar  Array, and Siberian Solar Radio Telescope, now named
Expanded Owens Valley Solar Array and Siberian Radioheliograph,
respectively) and general-purpose interferometers  (Very Large
Array) as well as the design of new interferometers, either
solar-dedicated like the Chinese Mingantu Ultrawide Spectral
Radioheliograph (MUSER) or for general astronomical use  (e.g. LOFAR
and ALMA) while ``first light'' from the Square Kilometre Array  (SKA)
is expected in the mid-2020s. The upgraded/new instrumentation
combined with the continuous operation of other important facilities
(e.g. Nan\c{c}ay Radioheliograph and RATAN-600) and  the efforts on
the modeling side promise exciting new results in the years to  come.

\section*{Author Contributions}

The manuscript was written by A.N. 

\section*{Acknowledgments}
I thank Dr. K.-L. Klein for his valuable comments on an earlier version of this
manuscript. I also thank Prof. C.E. Alissandrakis for his valuable comments on
the manuscript.

\bibliographystyle{frontiersinSCNS_ENG_HUMS}
\bibliography{incoherent_frontiers_accepted_arxiv}

\newpage

\section*{Figure captions}

\begin{figure}[h]
\begin{center}
\includegraphics[scale=.5]{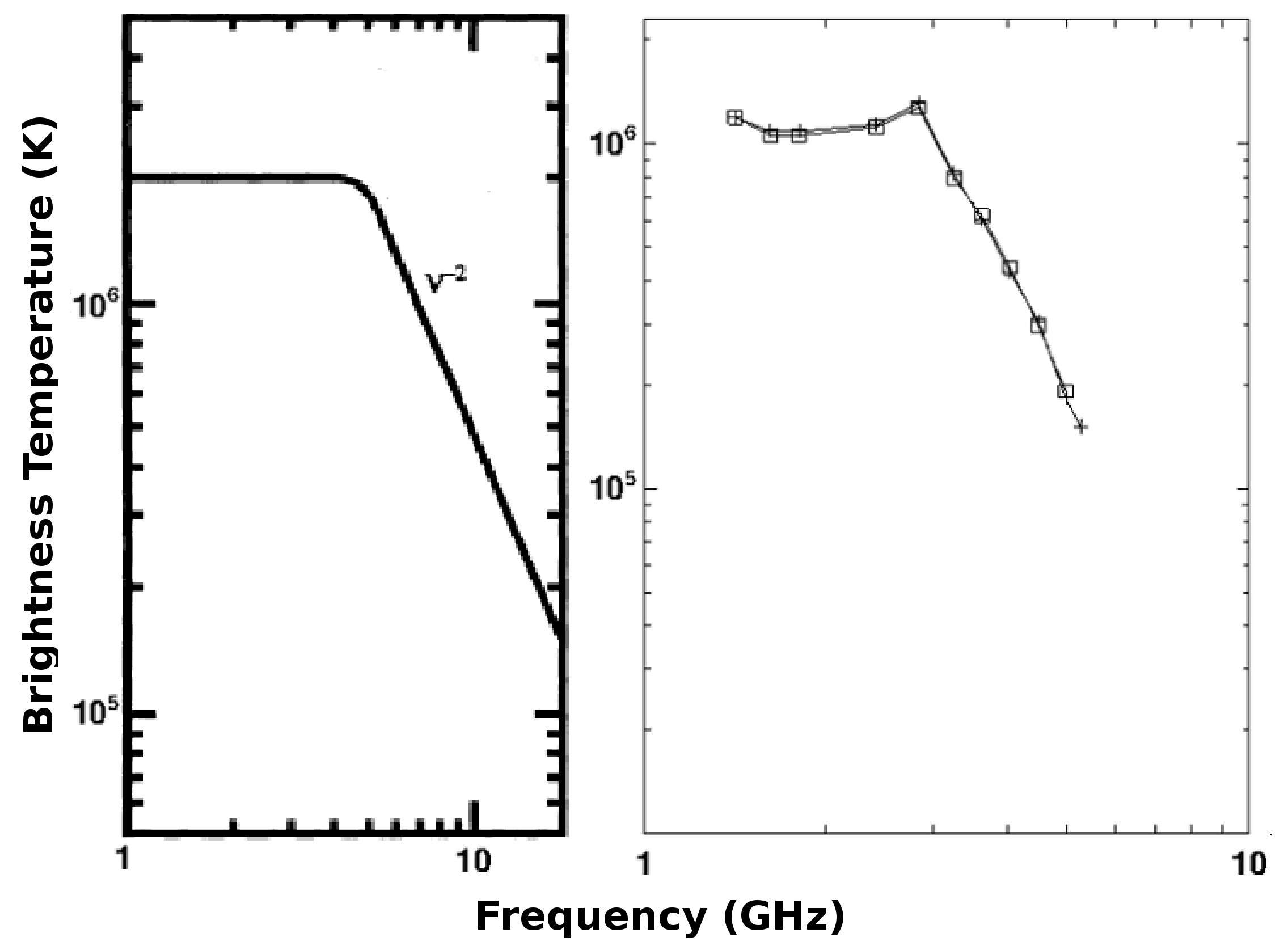}
\end{center}
\caption{Left panel: Model of the microwave brightness temperature spectrum
of free-free emission \citep[adapted from][]{Gary94}. \textcopyright{AAS}. 
Reproduced with permission. Right panel: Brightness temperature spectrum of 
part of an active region, with crosses and  boxes representing left-hand and
right-hand circular polarization,  respectively. The observations were
obtained with the OVSA at 16 frequencies in the range 1.4-8 GHz \citep[adapted 
from][]{Gary87}. \textcopyright{AAS}. Reproduced with permission}
\end{figure}

\begin{figure}
\begin{center}
\includegraphics[scale=.70]{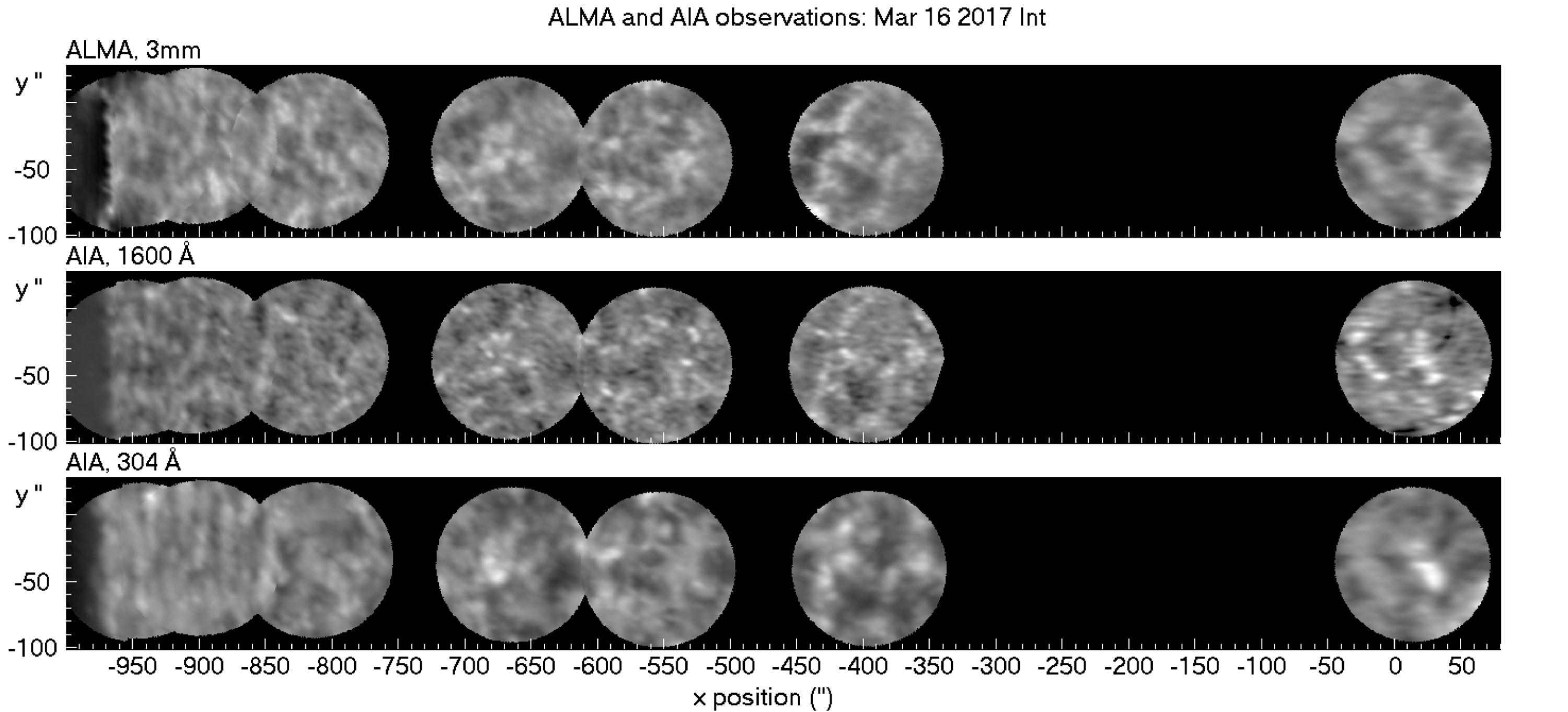}
\end{center}
\caption{The top row shows a composite (oriented in the south-east direction 
with respect to solar north) of seven 3 mm average ALMA images (each one 
computed from visibilities of duration 10 minutes) while the second and third 
rows show composites of AIA 1600 and 304 \AA\ average images for the same 
fields of view and time intervals as the ALMA images. The AIA images have been 
convolved with the ALMA beam \citep[from][]{Nindos18}. Reproduced with 
permission {\textcopyright}ESO.}
\end{figure}

\begin{figure}
\begin{center}
\includegraphics[scale=.70]{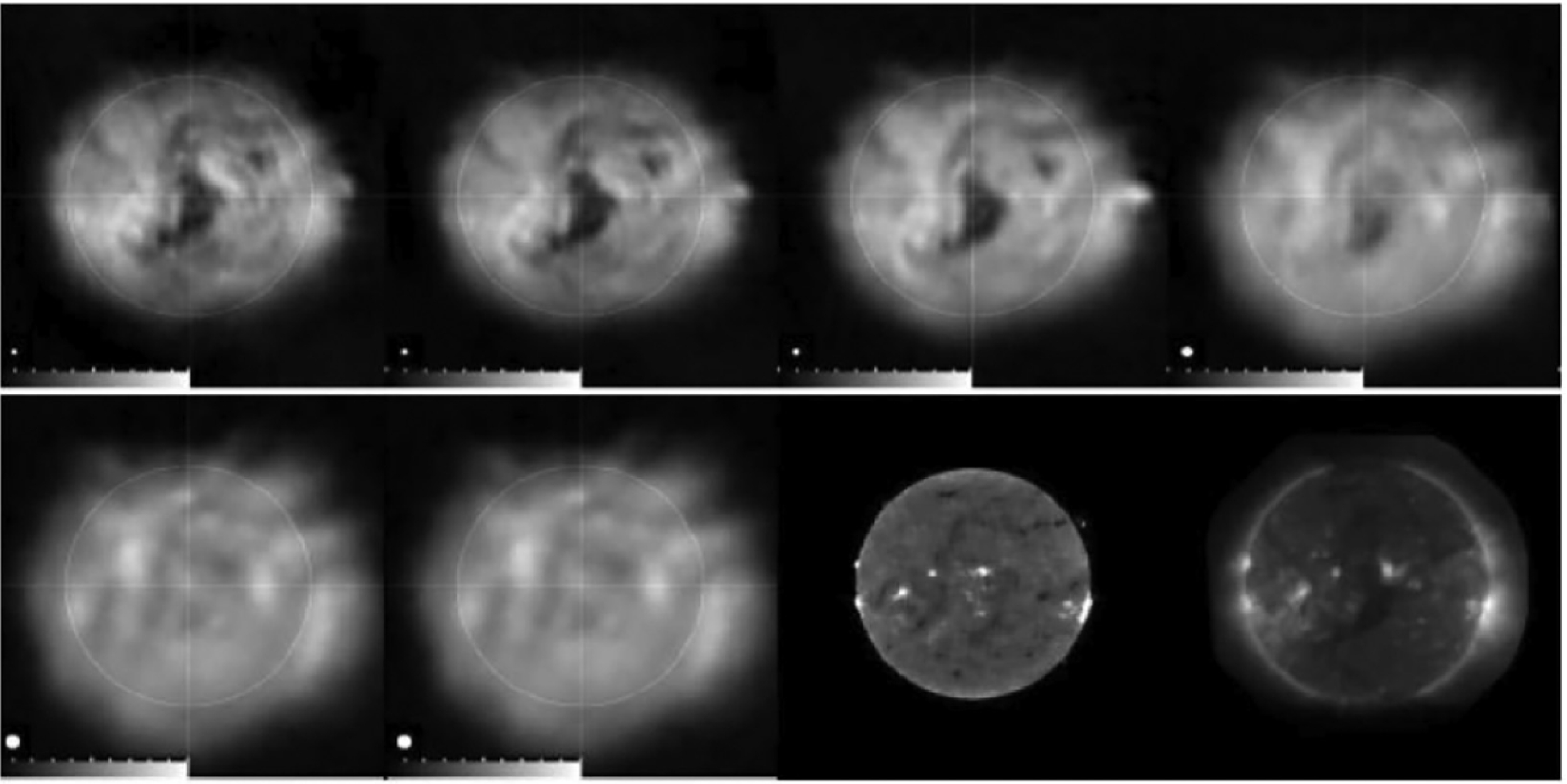}
\end{center}
\caption{Solar images on 2004 June 27. From left to right and from top to 
bottom: NRH images obtained at 432, 410, 327, 236, 164, and 150 MHz, NoRH image
at 17 GHz, and soft X-rays image obtained from the Solar X-Ray Imager (SXI) on 
board GOES12 \citep[from][]{Mercier09}. \textcopyright{AAS}. Reproduced with permission.}
\end{figure}

\begin{figure}
\begin{center}
\includegraphics[scale=.35]{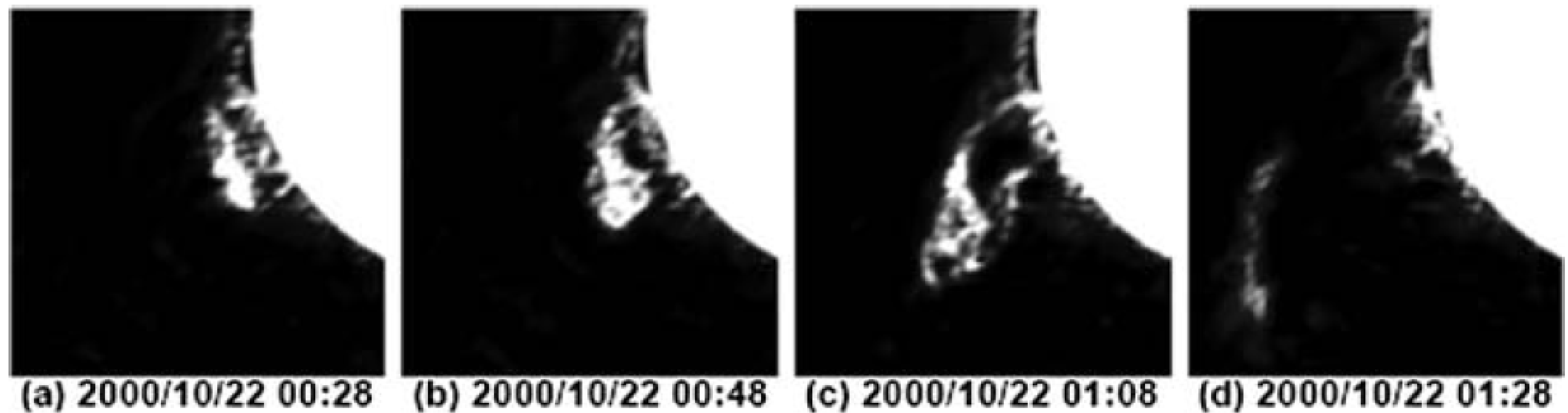}
\end{center}
\caption{17-GHz images of an eruptive prominence obtained by the NoRH 
\citep[from][]{Gopalswamy03}. \textcopyright{AAS}. Reproduced with permission.}
\end{figure}

\begin{figure}
\begin{center}
\includegraphics[scale=.40]{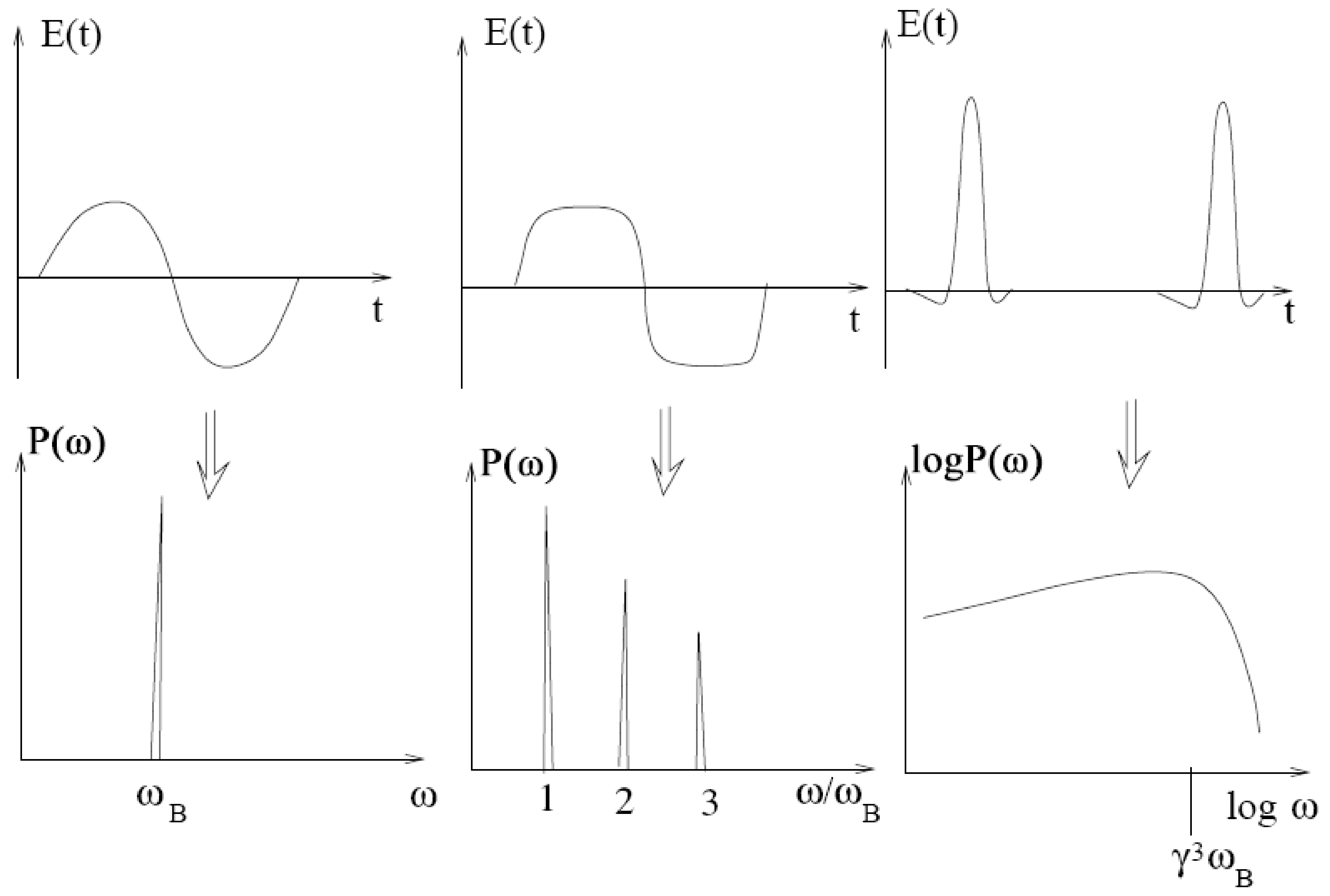}
\end{center}
\caption{Top row: Schematic time variability of the detected electric field 
that has been produced by a gyrating electron 
in the cases of small speed (left panel), thermal speed (middle panel) 
and relativistic speed (right panel). Bottom row: Schematic power spectra 
resulted from the cases of the top row.}
\end{figure}

\begin{figure}[t]
\begin{center}
\includegraphics[scale=.4]{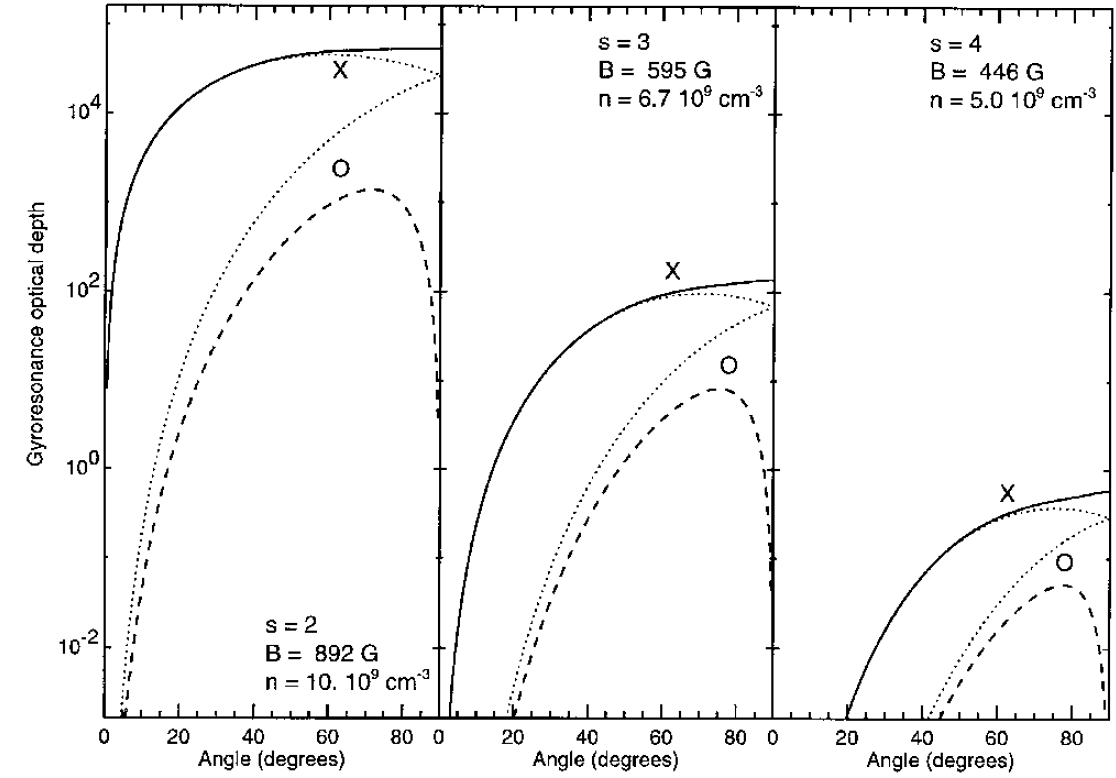}
\end{center}
\caption{Integrated optical depth, at 5 GHz, of the second, third, and fourth 
gyroresonance layers (left, middle, and right panels, respectively) as a 
function of the angle between the magnetic field and the line of sight. 
The scale height of the magnetic field is 10$^9$ cm, the electron temperature is
$3 \times 10^6$ K and the densities are denoted in the figure. The solid and
dashed lines correspond to $x$- and $o$-mode exact calculations, respectively. 
The dotted lines show $x$- and $o$-mode approximate calculations    
\citep[from][]{White97}. Reproduced with permission {\textcopyright}Springer 
Nature.}
\end{figure}

\begin{figure}[t]
\begin{center}
\includegraphics[scale=.75]{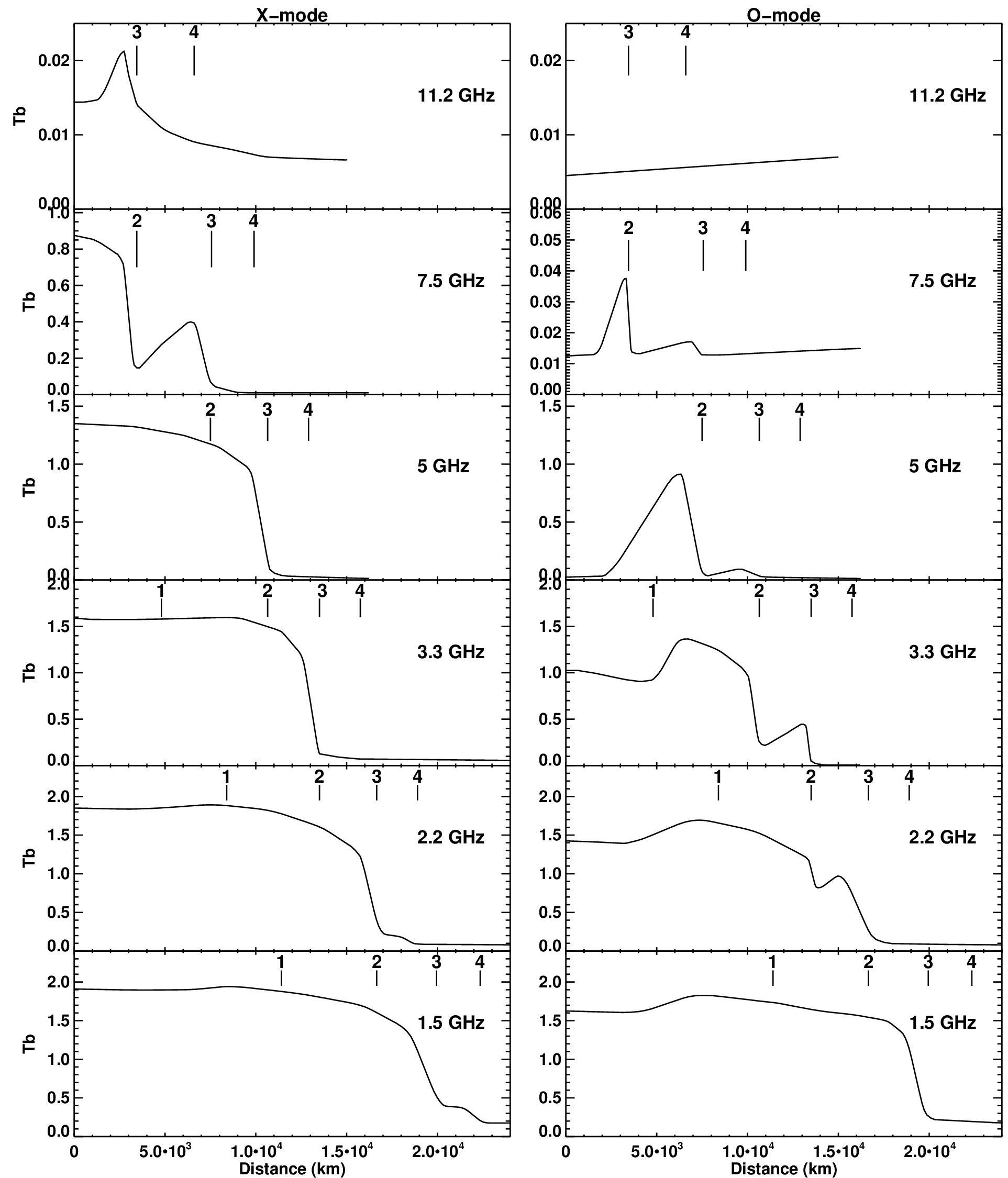}
\end{center}
\caption{Brightness temperature (in units of million degrees Kelvin) 
of the $x$-mode (left column) and $o$-mode (right panel) emission as a 
function of distance from the center of a dipole sunspot model at several 
frequencies (see text for details). The position at which the harmonic 
layers cross the base of the TR are marked with vertical straight lines.}
\end{figure}

\begin{figure}[t]
\begin{center}
\includegraphics[scale=.75]{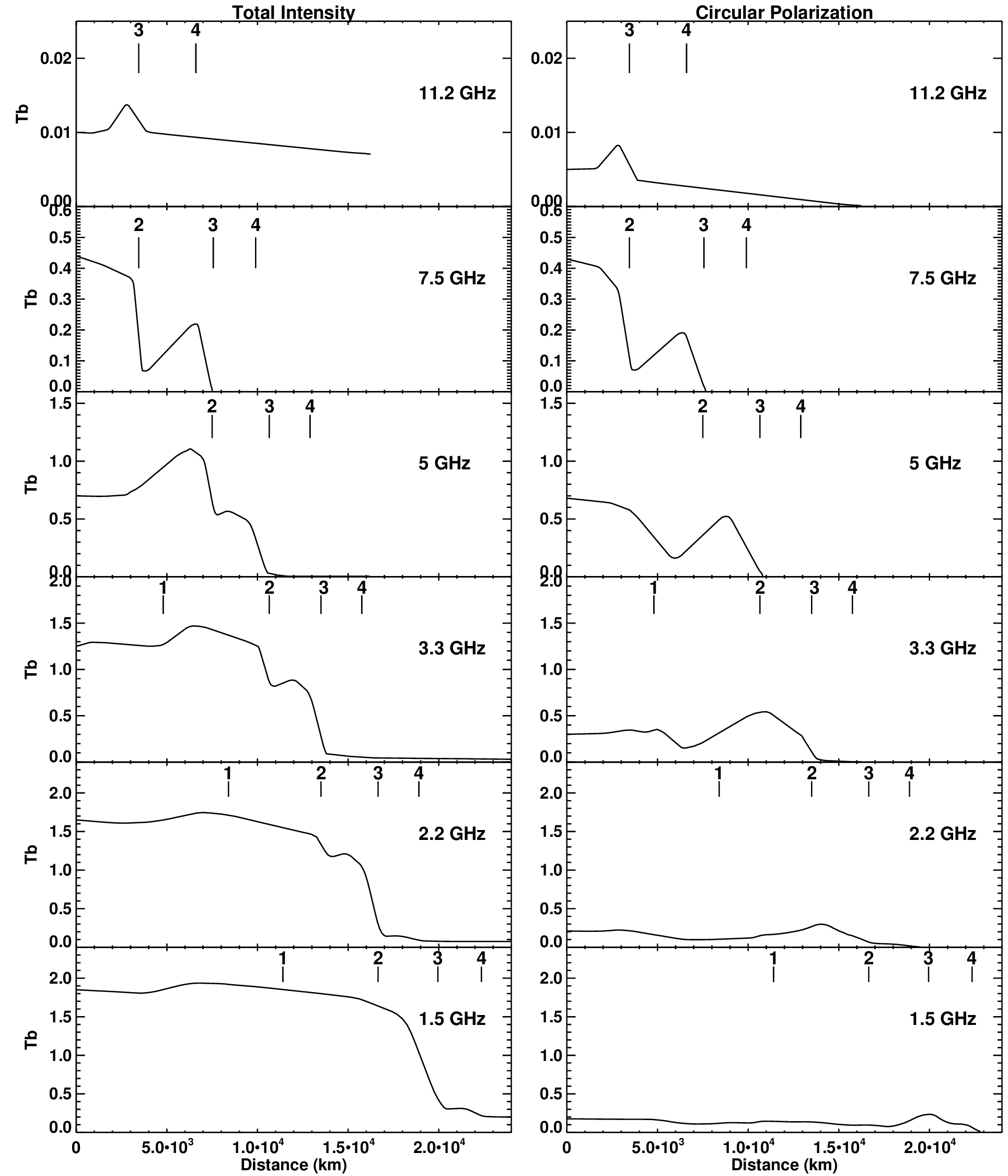}
\end{center}
\caption{Brightness temperature (in units of million degrees Kelvin) 
in $I$ (left column) and $V$ (right column) as a function of distance from 
the center of the dipole sunspot model of fig. 7, calculated from the radio 
models of fig. 7.}
\end{figure}

\begin{figure}
\begin{center}
\includegraphics[scale=.6]{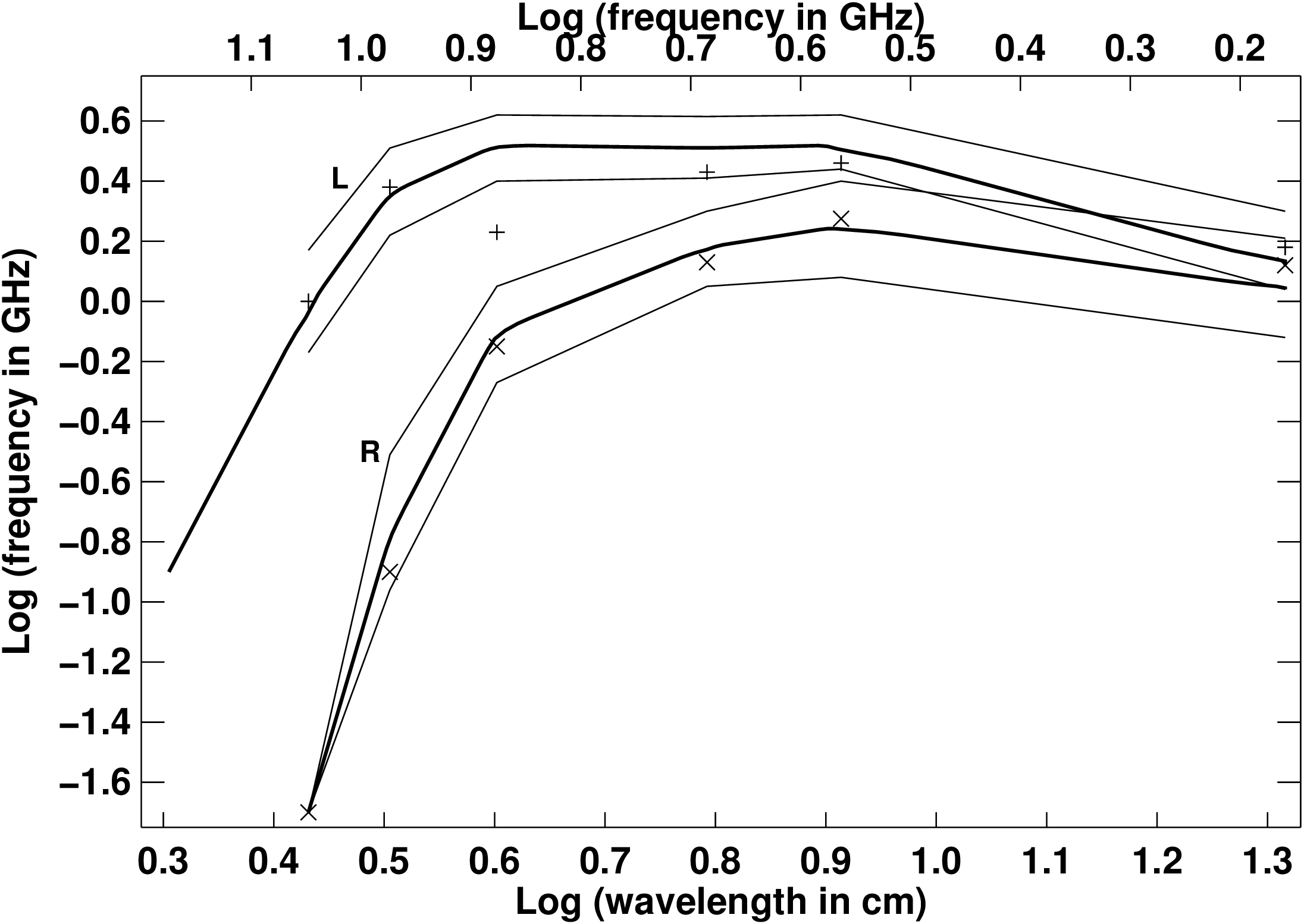}
\end{center}
\caption{Model flux density spectra in $L$ and $R$ polarizations ($x$- and
$o$-mode, respectively), computed from magnetic field
extrapolations for the cases: 
(i) $h_0=2000$ km, $F_c=6 \times 10^6$ erg cm$^{-2}$ s$^{-1}$, (thick lines), 
(ii) $h_0=2000$ km, $F_c=3 \times 10^6$ erg cm$^{-2}$ s$^{-1}$,  (iii) $h_0=2000$
km, $F_c=1.2 \times 10^7$ erg cm$^{-2}$ s$^{-1}$. In case (i) the flux was
computed at frequencies from 2 to 15 GHz while in cases (ii) and (iii) it
was computed for 1.5, 3.7, 4.8, 7.5, 9.4, and 11.1 GHz. The observed
flux density spectra in $R$ ($\times$) and $L$ ($+$) are also given 
\citep[from][]{Nindos96}. Reproduced with permission {\textcopyright}Springer 
Nature.}
\end{figure}

\begin{figure}[t!]
\begin{center}
\includegraphics[scale=.6]{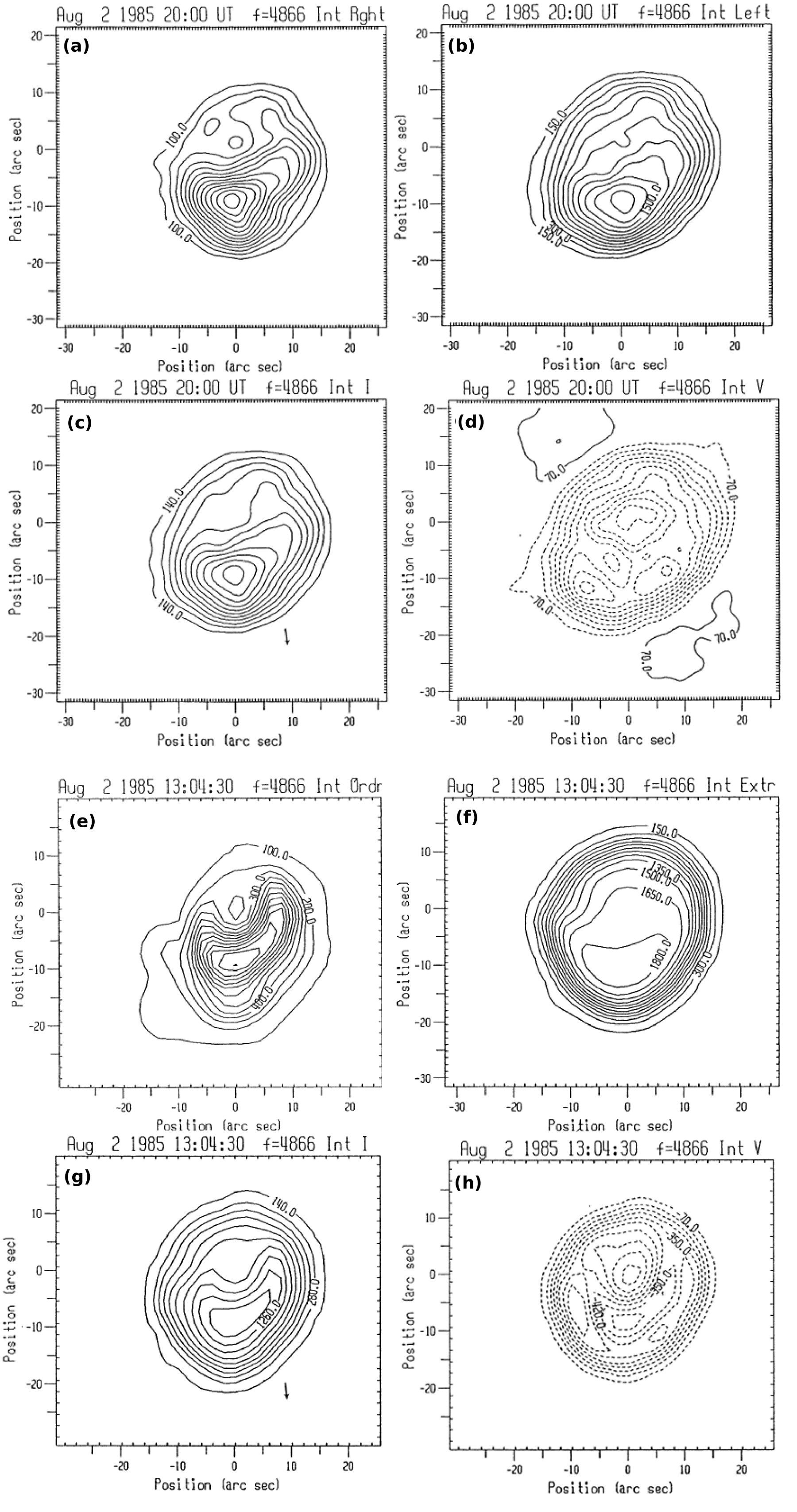}
\end{center}
\caption{Maps of active region 4862 observed with the VLA at 5 GHz in
Stokes $I$, $V$ (panels c and d, respectively)  and in $R$ and $L$  
polarization (panels a and b, respectively). The contours are in  brightness 
temperature steps of $1.4 \times 10^5$, $0.7 \times 10^5$, $1.5 \times 10^5$, 
and 10$^5$ K, respectively. Dashed contours indicate negative values. Panels 
(e) and (f) show models of
active region 4862 at 5 GHz in $R$ and $L$ which were computed with
$\alpha=-2.4 \times  10^{-5}$ km$^{-1}$ and $\alpha=2.4 \times 10^{-5}$ km$^{-1}$,
respectively. Panels (g) and (h) show model $I$ and $V$ maps, respectively,
calculated from the models of panels (e) and (f). The arrow in panels (c) and
(g) shows the direction of the limb \citep[adapted from][]{Nindos96}. 
Reproduced with permission {\textcopyright}Springer Nature.}
\end{figure}

\begin{figure}
\begin{center}
\includegraphics[scale=.6]{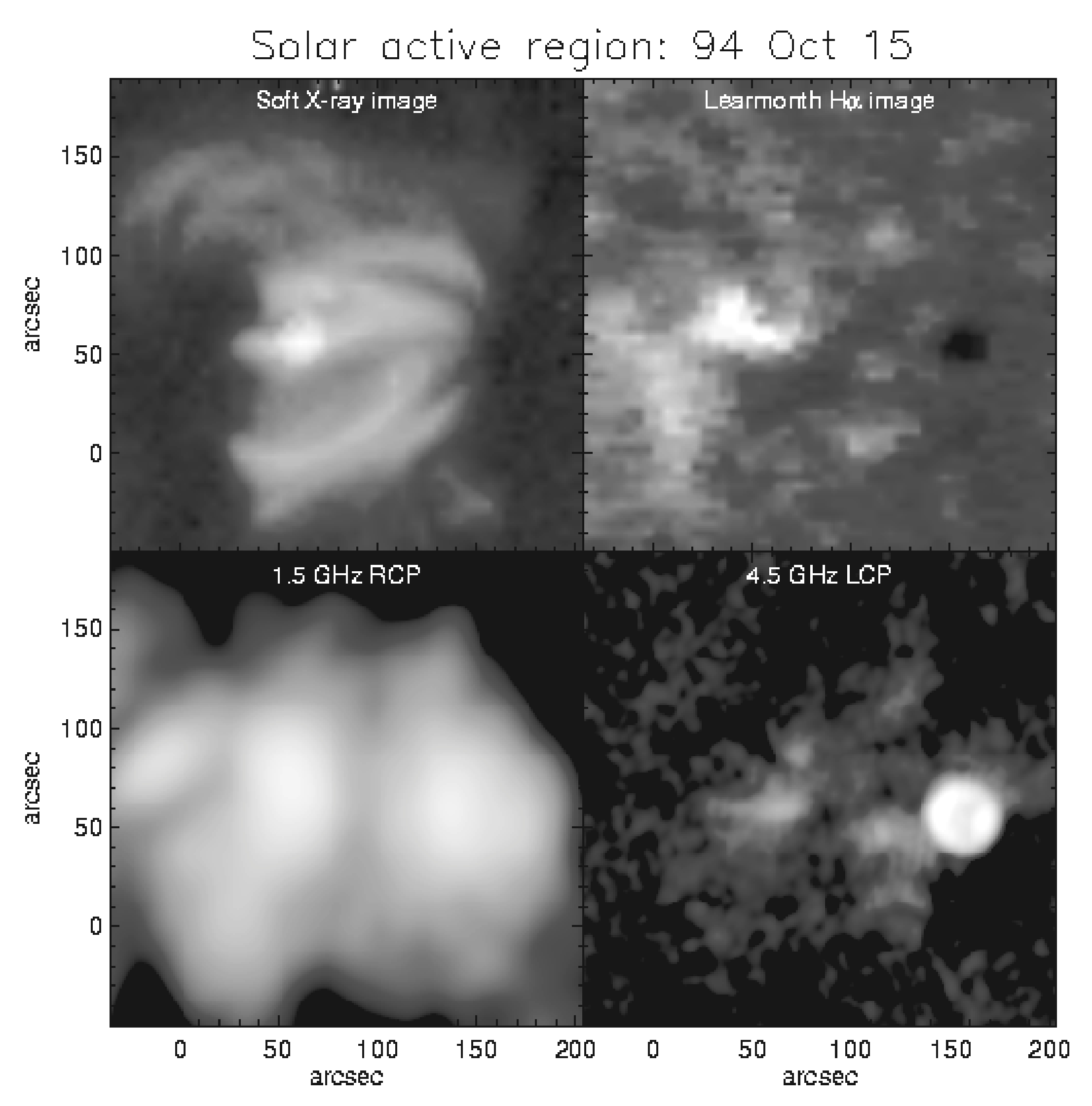}
\end{center}
\caption{Images of an active region. Top left: soft X-ray image obtained 
by Yohkoh Soft X-ray Telescope (SXT). Top right: H$\alpha$ image from Learmonth
Observatory. Bottom left: 1.5 GHz VLA image in  right circular polarization. 
Bottom right: 4.5 GHz VLA image in left circular polarization (image credit: 
S.M. White).}
\end{figure}

\begin{figure}
\begin{center}
\includegraphics[scale=.6]{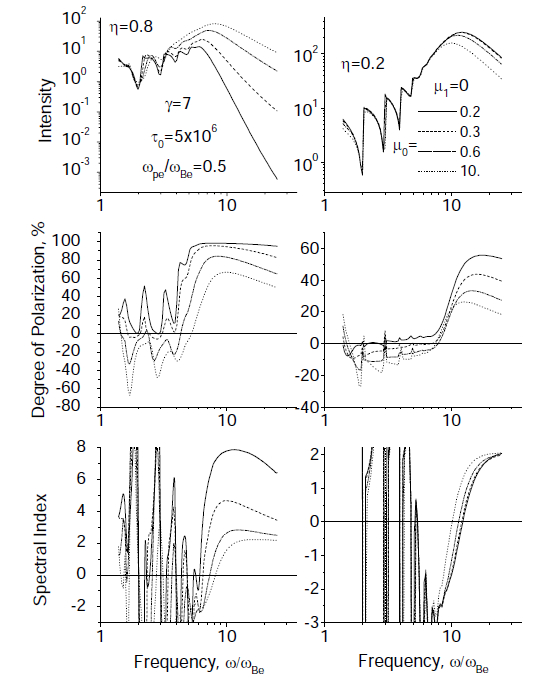}
\end{center}
\caption{Gyrosynchrotron intensity, degree of polarization and spectral index
as a function of frequency for various values of $\mu_0$ in the Gaussian
loss-cone type pitch-angle distribution $f \propto \exp [-\mu^2/\mu_0^2]$.
The computations presented in the left and right columns have been made for 
$\eta = 0.8$ and 0.2, respectively, where $\eta$ is the cosine of the angle 
between the magnetic field and the line of sight \citep[from][]{Fleishman03b}.
\textcopyright{AAS}. Reproduced with permission.}
\end{figure}

\begin{figure}
\begin{center}
\includegraphics[scale=.70]{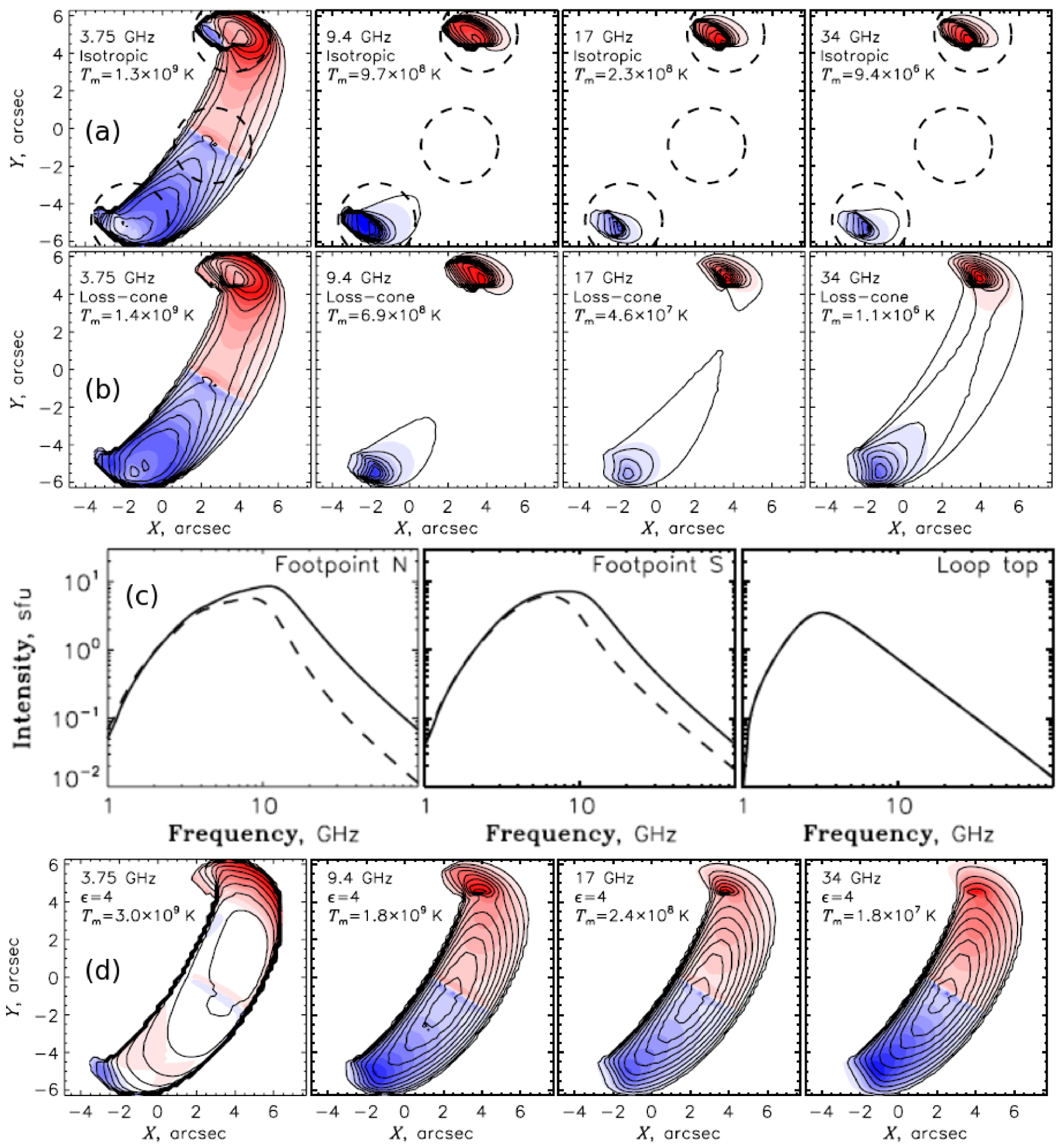}
\end{center}
\caption{Brightness temperature maps of the gyrosynchrotron emission
from a model loop for an isotropic electron pitch-angle distribution
[row (a)], and a loss-cone pitch-angle distribution [row (b)]. In both
cases the density of the accelerated electrons is constant along the
loop. Row (c): Flux density spectra of the northern footpoint source
(left), southern footpoint source (middle) and loop top source (right)
resulted from the models of rows (a) and (b) (solid lines and dashed
lines, respectively). These spectra were computed in the areas defined
by the dashed circles of row (a). Row (d): same as row (b) but
with an inhomogeneous spatial profile of the electron density along
the loop.  In rows (a), (b), and (d), the contours denote
intensities evenly distributed from zero to the maximum brightness
temperature which is given in each panel. The red and blue
colors denote the circular polarization (right and left sense,
respectively). Adapted from \cite{Kuznetsov11}. \textcopyright{AAS}. 
Reproduced with permission.}
\end{figure}

\begin{figure}
\begin{center}
\includegraphics[scale=.84]{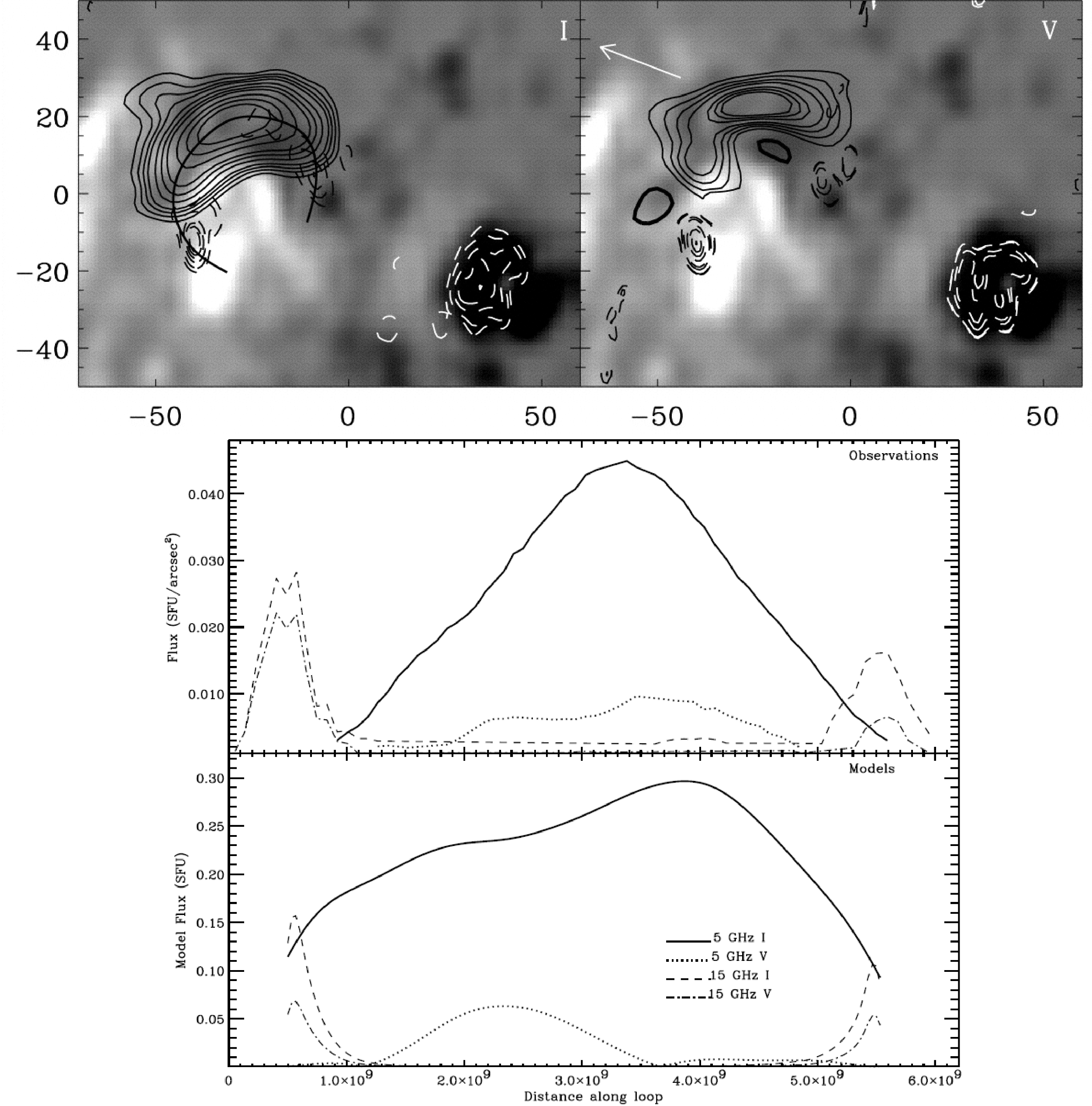}
\end{center}
\caption{Top row: The 1992 July 1 flare. Contour plots of the flare radio 
emission maximum observed  by the VLA. The gray-scale background is a
photospheric magnetogram. The $I$ maps are on the left, and the $V$
maps on the right.  The solid and dashed contours show 5 GHz and
15 GHz emission, respectively. In the 5 GHz $V$ map, the thick
contours represent positive brightness temperatures. In both images the white
contours show emission from the sunspot at 15 GHz. The  arrow shows the 
direction of the limb. Middle row: One-dimensional profiles of the flare 
computed along the black curve of the top left panel. Bottom row: spatial 
profiles of the gyrosynchrotron models  (see text for details) as a function 
of the distance along the loop. For comparison with the  observations, the 
profiles have been computed after the models were convolved with the 
appropriate VLA beam. In both panels the absolute values of the $V$ profiles 
are presented \citep[adapted from][]{Nindos00}. \textcopyright{AAS}. Reproduced
with permission.}
\end{figure}

\begin{figure}
\begin{center}
\includegraphics[scale=.70]{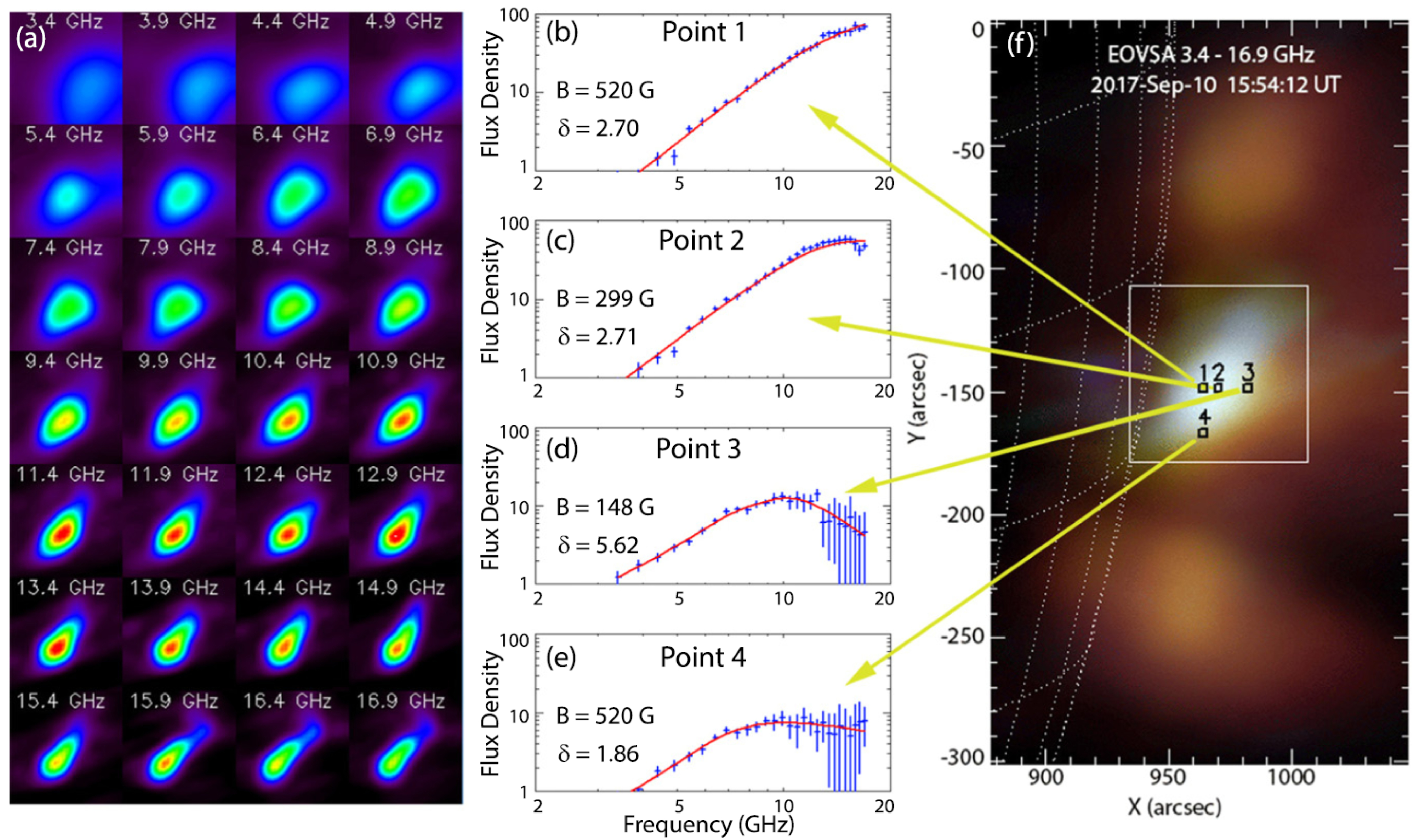}
\end{center}
\caption{(a) Images of the microwave emission from a partially occulted 
limb flare obtained at 28 frequencies with EOVSA. The field of view of the
images corresponds to the white box of panel (f). (b)-(e) Flux density
spectra in those pixels of the images of panel (a) that correspond to points
1-4 of panel (f), together with model spectral fits (red lines). (f) True-color
display of the EOVSA dataset after combining images at the 28 frequencies that
are marked in panel (a). From \cite{Gary18}. \textcopyright{AAS}. Reproduced 
with permission.}  
\end{figure}

\end{document}